\documentclass[aip, amsmath, amssymb, reprint]{revtex4-2}
\usepackage{graphicx} 
\usepackage{bm} 
\usepackage[utf8]{inputenc}
\usepackage[all]{nowidow}
\usepackage{etoolbox}
\makeatletter
\def\@email#1#2{
 \endgroup
 \patchcmd{\titleblock@produce}
  {\frontmatter@RRAPformat}
  {\frontmatter@RRAPformat{\produce@RRAP{*#1\href{mailto:#2}{#2}}}\frontmatter@RRAPformat}
  {}{}
}
\makeatother

\begin{document}

\title{Data-driven approach for benchmarking DFTB-approximate excited state methods}

\author{Andrés I. Bertoni}
\author{Cristián G. Sánchez$^{*}$}
\email[Corresponding author.~E-mail:~]{csanchez@mendoza-conicet.gob.ar}
\affiliation{Instituto Interdisciplinario de Ciencias Básicas (ICB-CONICET), Universidad Nacional de Cuyo, Padre Jorge Contreras 1300, Mendoza 5502, Argentina}

\begin{abstract}
In this work we propose a chemically-informed data-driven approach to benchmark the approximate density-functional tight-binding (DFTB) excited state (ES) methods that are currently available within the \texttt{DFTB+} suite. By taking advantage of the large volume of low-detail ES-data in the machine learning (ML) dataset, QM8, we were able to extract valuable insights regarding the limitations of the benchmarked methods in terms of the approximations made to the parent formalism, density-functional theory (DFT), while providing recommendations on how to overcome them. For this benchmark, we compared the first singlet-singlet vertical excitation energies ($E_1$) predicted by the DFTB-approximate methods with predictions of less approximate methods from the reference ML-dataset. For the nearly 21,800 organic molecules in the GDB-8 chemical space, we were able to identify clear trends in the $E_1$ prediction error distributions, with respect to second-order approximate coupled cluster (CC2), showing a strong dependence on chemical identity.
\end{abstract}
 
\maketitle

\section{Introduction}
Machine learning (ML) techniques are becoming ubiquitous in all areas of scientific research, and quantum chemistry is no exception. There are efforts in this area of research that are directed toward the automated prediction of molecular electronic properties from structure \cite{Fabrizio2019, Yang2020, Joung2021, Choi2022}. In particular, the prediction of excited state (ES) properties, such as photo-absorption excitation energies, could enable the data-driven discovery of useful chromophores, such as organic dyes with potential uses in medical imaging and optoelectronic devices. Advances in ML chemical modelling are accompanied by the creation of datasets of increasing size and quality, containing chemical information from experiments and theoretical calculations \cite{Ramakrishnan2015, Kayastha2022}. These large collections of curated data are generally assembled to train deep learning models that are capable of accurately predicting the electronic properties of other molecules within the same chemical subspace as the training set. Although these datasets currently provide a limited level of detail, they stand out because they span a large volume of the chemical space.

In this work we recognise additional value in ML-datasets and propose a chemically-informed data-driven approach to benchmark the approximate density-functional tight-binding (DFTB) excited state methods that are currently available in the \texttt{DFTB+} suite. For this benchmark, we compared the first singlet-singlet vertical excitation energies ($E_1$) predicted by the approximate methods with less approximate methods from the reference ML-dataset. For the nearly 21,800 organic molecules in the GDB-8 chemical space, we were able to identify clear trends in the $E_1$ prediction error distributions, with respect to second-order approximate coupled cluster (CC2), with a strong dependence on chemical identity. In other words, we took advantage of the large volume of low-detail ES-data in the QM8 ML-dataset, and combined it with molecular descriptors based on $\pi$-conjugation and chemical identity, in order to extract valuable insights regarding the limitations of the benchmarked methods in terms of the approximations made to the parent formalism, density-functional theory (DFT). As this benchmark was performed for a large number of molecules with a diverse representation of chemical groups, it serves as a comprehensive complement to other reported benchmarks concerning specific implementations in \texttt{DFTB+} \cite{Niehaus2009, Trani2011, Dominguez2013, Kranz2017, Fihey2019}.

These acquired insights allowed us to provide recommendations to the \texttt{DFTB+} user and developer communities on the limitations that may be encountered and how to overcome them. As a result of this benchmark, we provide (i) a ``rule of thumb'' for computing $E_1$ with the ES-methods and corrections currently implemented in \texttt{DFTB+}, (ii) a proof-of-concept \footnote[3]{To be used for ES calculations with fixed nuclei only, since we have not performed a re-parameterization of the repulsion splines.} minimally polarized version of a popular SK-parameter set, (iii) recommendations for next steps in \texttt{DFTB+} development, and (iv) an extension of the QM8 ML-dataset with DFTB-results.

\section{Overview of methods}
\texttt{DFTB+} is a popular and open source scientific software package that offers a fast and scalable implementation of the density functional based tight binding (DFTB) method \cite{Elstner1998}. Among the many implemented features, which are being actively developed \cite{Hourahine2020} by its vast community of users, there is available a set of methods for simulating the photo-absorption spectroscopy of molecular systems; namely, the time-dependent DFTB method (TD-DFTB) and the DFTB-approximate particle-particle random phase approximation (ppRPA) \cite{Yang2017}. In \texttt{DFTB+}, TD-DFTB exists as two distinct but equivalent implementations: one is based on Casida's linear response (LR) approach \cite{Niehaus2001} and the other is a real-time implementation based on the semi-classical Ehrenfest method for nuclear-electronic dynamics \cite{Bonafe2020}.

Being a semi-empirical method, DFTB is a considerably more approximate method than its parent \textit{ab initio} method, DFT \cite{Hohenberg1964, Kohn1965}. However, these approximations are precisely the reason for its superior computational performance. The relatively lower accuracy of DFTB is widely compensated by a substantial drop in computational cost, which extends the feasibility upper limit in size, timescale and complexity of the systems to be studied. As a point of comparison, TD-DFTB approximations result in a computational scaling reduced by three orders of magnitude with respect to TD-DFT in its default formulation (from $N^{6}$ to $N^{3}$, where $N$ is the number of atoms in the system) \cite{Hourahine2020}. In other words, the approximations of DFTB bring quantum mechanics to systems that otherwise would be treated with classical force fields. As a result, DFTB is being widely employed in pure and hybrid QM-schemes for the simulation of the spectroscopic properties of nanostructures \cite{Berdakin2022}, photoactive materials \cite{Lien-Medrano2022} and complex biological systems \cite{Maity2020} of many thousands of atoms.

Our work is aimed to both experts and beginners in the DFTB+ community that share an interest in the computation of excitation energies within the DFTB framework. Thus, we have decided to include into the electronic supplementary material (ESI) brief overviews of the methods revised in this work, which will aid the reader in the discussion of the results. However, we strongly recommend the readers to refer to the original publications for the respective implementations of DFTB \cite{Hourahine2020}, TD-DFTB (DFTB-Casida) \cite{Niehaus2009} and pp-DFTB (DFTB-ppRPA) \cite{Yang2017}. We also want to recommend a pedagogical introduction to those who are completely new to DFTB \cite{Koskinen2009}.

\section{Computational Details}
\subsection{Reference dataset}
The QM8 reference dataset \cite{Ramakrishnan2015} was created for the development of machine learning models. Although this ML-dataset does not provide high-detail data for the low-lying electronic transitions (e.g., corresponding transition electric dipole moments or excited state symmetries), it instead provides a large amount of low-detail ES quantities, calculated with CC2 and linear-response time-dependent DFT (TD-DFT): it comprises the first two singlet-singlet vertical excitation energies ($E_1$ and $E_2$) and oscillator strengths ($f_1$ and $f_2$) for the 21786~organic molecules, with up to 8~CONF atoms (up to 26~CHONF atoms), which make the GDB-8 subset of the GDB-9 chemical space \cite{Ramakrishnan2014}.

As Ramakrishnan \textit{et~al.}, we have decided to take CC2 with the polarized triple-zeta valence second-generation basis set (def2TZVP) as the reference method. The authors based their decision on the fact that CC2/def2TZVP was shown to predict valence excitation energies in very close proximity to experimental data \cite{Send2011} and to a higher-order computational prediction \cite{Kannar2014}, with mean absolute errors of 0.12~eV and 0.10~eV, respectively.

\subsection{Computing the first singlet vertical excitation energy}
Prior to the computation of excitation energies, the molecular geometries were re-optimised with SCC-DFTB and a chosen parameter set from \texttt{www.dftb.org}. Geometries were relaxed via the limited-memory Broyden-Fletcher-Goldfarb-Shanno (L-BFGS) algorithm, in order to achieve maximum force components below $10^{-5} \, E_h/a_0$. For each DFTB-optimised geometry within the addressed chemical sub-space, we computed $E_1$ with the two approximate methods available for this task in the last version of \texttt{DFTB+} \cite{Hourahine2020}: DFTB-Casida and DFTB-ppRPA. Currently, these DFTB-approximate excited state methods are only implemented at the SCC-DFTB level of theory, for molecular systems. Although we were only interested in $E_1$, the number of computed excitation energies was set up to~20, when possible, to aid the convergence of the eigenvalue solvers. Specifically for DFTB-ppRPA calculations, we also set the maximum number of virtual states to~1000 and provided the Hubbard-like parameters from the Appendix~K of the \texttt{DFTB+} manual \cite{DFTBManual}. For both approaches, we employed the general purpose parameter set 3OB(-3-1) \cite{Gaus2013}, which has parameters for a wide variety of elements present in biological and organic systems. For 3OB, the Hamiltonian and overlap matrix elements were pre-computed using DFT with the Perdew-Burke-Ernzerhof (PBE) functional.

In this work, we decided to compare the error of each DFTB-approximate method in the prediction of $E_1$, with respect to CC2/def2TZVP ($\Delta_{CC2} E_1$), with results from the parent method, DFT. With this analysis, our aim is to detect which of the limitations in the prediction of $E_1$ are being shared between the related methods and which are unique to the standard formulation of DFTB. As the parent method for this comparison we selected TD-DFT with the hybrid HF-XC variant of the local PBE functional (PBE0) and the polarized split-valence second-generation basis set (def2SVP). This choice corresponds to the method within QM8 that is most related to DFTB: a DFT method using the most local XC functional and the smallest basis set. Despite the hybrid functional PBE0 is the closest one in the ML-dataset to PBE (i.e., the GGA functional employed to calculate the electronic parameters for DFTB), systematic differences between local and hybrid functionals should be expected regarding electronic excitations within TD-DFT; this fact will contribute to the overall differences in performance between TD-DFT and TD-DFTB, particularly for $\pi$-type excitations. For more information on this topic, we would like to recommend the work of Maier and Kaupp \textit{et~al.} \cite{Maier2016}, which includes a systematic comparison between (semi-)local and global hybrid functionals, with respect to the prediction of TD-DFT electronic excitation energies, for a small set of organic molecules. Values of $E_1$ from the reference method, CC2/def2TZVP, and the time-dependent extension of the parent method, TD-DFT:PBE0/def2SVP, were directly taken from QM8; these values were originally computed from geometries relaxed via DFT with the Becke 3-parameter Lee-Yang-Parr hybrid XC-functional (DFT/B3LYP) and the polarized Pople-style 6-31G(2df,p) basis set.

To address the limitations encountered, we also included extensions to the standard SCC-DFTB method in the calculation of $E_1$, whose implementations are available in \texttt{DFTB+}. Specifically, we included long-range corrections (LC), on-site corrections (OC), and a proof-of-concept custom extension of the SK-parameter set to include minimal polarization on H atoms only (3OB(H*)); see the ESI for the theoretical and technical details of these extensions.

\subsection{Using molecular descriptors for data analysis}
The molecules in the GDB-8 chemical subspace were generated following synthetic feasibility rules, but the combinatorial nature of the structure generation algorithm may have resulted in structures that may be considered unusual. However, we classified the molecules into chemical families based on their functional group with the highest level of conjugation, because we expect the functional group with the largest $\pi$-conjugation in a molecule to be the main contributor to the frontier molecular orbitals, thus characterising the lowest electronic excitation. Under this premise, the rest of the molecule would be acting as a local chemical environment that modulates the excitation energy of the main functional group, regardless of how rare the chemical arrangement is. This way, the conclusions to be drawn may also apply to molecules larger than those in the dataset (i.e., of more than 26~atoms), if the additional atoms do not extend the conjugated systems; however, if that is the case, then these larger molecules would have their $E_1$ closer to the visible light region, but would still be members of the $\pi$-conjugated family, and thus would be expected to suffer from the same limitations as the ones we will discuss in this work.

For each molecule in the ML-dataset, we used the Python library \texttt{rdkit} \cite{RDKit} to identify its level of $\pi$-conjugation and the presence of certain chemical groups, through the analysis of their SMILES strings \cite{Weininger1988}. In that manner, molecules were assigned to one of three main families of chemical compounds: (i) $\pi$-conjugated unsaturated molecules, (ii) non-conjugated unsaturated molecules (i.e with isolated double or triple bonds) and (iii) saturated molecules (i.e., with only single bonds). Within the family of $\pi$-conjugated unsaturated molecules, we made a distinction between molecules with conjugated groups of 3~atoms and molecules with larger conjugated groups (i.e., with more than 3~atoms). Via \texttt{rdkit} tools for handling the SMILES arbitrary target specification (SMARTS) language, we assigned a more detailed chemical-identity to the molecules of each principal chemical family. For this task, extra chemical descriptors were employed in a hierarchical fashion (i.e., based on the order of assignation of labels); in other words, a molecule assigned to a particular chemical sub-family was removed from the pool of molecules to be identified as part of any other sub-group within the same main family. The SMARTS fragments employed are available in Table~I of the ESI.

\section{Results and Discussion}

\subsection{Initial diagnostics}
\begin{figure*}
\includegraphics[scale=0.46]{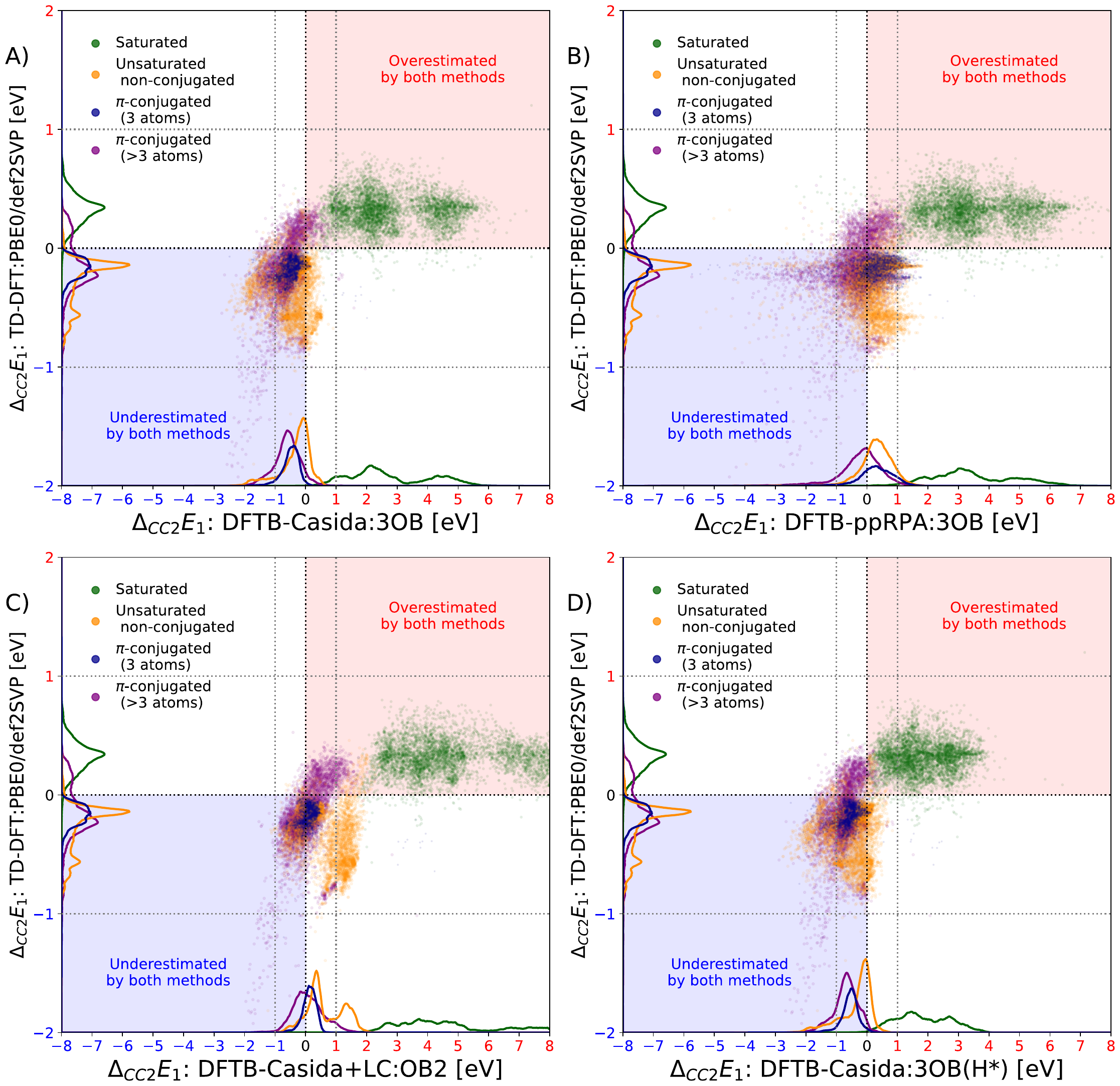}
\caption{\label{Figure1} Comparison of prediction errors $\Delta_{CC2} E_1$ for TD-DFT:PBE0/def2SVP and four different DFTB-approximate approaches: (A) Casida:3OB, (B) ppRPA:3OB, (C) Casida+LC:OB2 and (D) Casida:3OB(H*). The scattered datapoints correspond to each of the nearly 21,800 molecules in the GDB-8 chemical subspace. The datapoints and the projected histograms were coloured according to the main chemical identity of the compounds: green for saturated molecules, orange for non-conjugated molecules with an isolated double or triple bond, and blue and purple for $\pi$-conjugated molecules with~3 or more conjugated atoms, respectively. See the first column of Tables~IV~\&~V, in the ESI, for the number of occurrences per chemical family and subgroup.}
\end{figure*}

The prediction accuracy of $E_1$ was compared between the approximate DFTB methods and TD-DFT:PBE0/def2SVP. Specifically, we started with the comparison of prediction errors $\Delta_{CC2} E_1$ for DFTB-Casida:3OB and DFTB-ppRPA:3OB, which are shown as datapoints in panels $A$ and $B$ of Fig.~\ref{Figure1}, respectively. The scattered points correspond to each of the nearly 21,800 molecules in the GDB-8 chemical subspace, and are coloured to reflect their membership to one of the main chemical families we defined before. Datapoints were located on the 2D planes according to the prediction error $\Delta_{CC2} E_1$, as computed with the benchmarked method (horizontal axis) and TD-DFT:PBE0/def2SVP (vertical axis). Accordingly, molecules in the lower-left quadrant (blue background) have $E_1$ values that are \textit{underestimated} by both methods, while those in the upper-right quadrant (red background) have their $E_1$ \textit{overestimated} by both methods. In addition, each group of coloured points was also projected onto each axis, as histograms, to show the $E_1$ error distribution for each chemical family, for each of the compared methods.

First of all, Fig.~\ref{Figure1} shows that the error of $E_1$ predictions strongly depends on the molecules' main chemical identity. Based on error distributions of $E_1$ for TD-DFT, Ramakrishnan \textit{et~al.} first noticed that predictions could be divided into two distinctive sets of molecules: saturated and unsaturated; the authors observed that, overall, $E_1$ was underestimated for unsaturated molecules and overestimated for saturated molecules. In panel~A of Fig.~\ref{Figure1}, we found broadly the same saturated-unsaturated pattern for DFTB-Casida:3OB, as the vast majority of points are either in the underestimation or overestimation quadrants. This reproduction of general trends between TD-DFTB and TD-DFT is not entirely surprising, as the former method is an approximation to the latter.

On the one hand, molecules with extended $\pi$-conjugation are expected to exhibit low-lying excitations of the $\pi$-type ($\pi \rightarrow \pi^{*}$ or $n \rightarrow \pi^{*}$). Therefore, DFTB-Casida:3OB systematic underestimation of $E_1$ for unsaturated molecules should be arising from the DFT-inherited self-interaction error (SIE). This well-known error involves individual electrons interacting with their own density \cite{Lundberg2005}. The SIE is related to the local approximation of the XC contribution to the charge interaction energy, and arises from the Coulomb and exchange terms incorrectly not cancelling each other exactly at large distances \cite{Perdew1981}. As the solutions of SCC-DFTB are found self-consistently \cite{Lundberg2011}, the SIE leads to an incorrect charge distribution in the ground state, with the consequent artificial stabilisation of the delocalised solutions \cite{Savin1996, DellaSala2002, Kronik2014}. This results in the underestimation of HOMO-LUMO gaps \cite{Nishimoto2016} and excitation energies or, equivalently, in the overestimation of dipole polarizabilities \cite{Grimme2003}. This is why TD-DFT with local or semi-local functionals is not recommended for low-lying electronic transitions involving $\pi$-conjugated states \cite{Champagne1998, Champagne2000, Grimme2003}, nor it is recommended for charge-transfer (CT) excitations \cite{Maier2016, Dreuw2003, Dreuw2005}. The SIE is particularly problematic for CT-excitations \cite{Tozer2003}, most notably for systems with excess charge (e.g., ions \cite{Grozema2009} or radical ions \cite{Chai2008, Kubar2010}) and for transitions involving occupied and virtual molecular orbitals (MOs) with minimal spatial overlap \cite{Peach2008}.

On the other hand, saturated molecules are expected to exhibit mainly $\sigma$-type excitations ($\sigma \rightarrow \sigma^{*}$ or $n \rightarrow \sigma^{*}$). In some cases, Rydberg excitations may appear as the lowest-lying electronic excitations; however, without the use of a very diffuse basis set, Rydberg states should be outside the scope of both DFT and DFTB methods. The reasons behind DFTB-Casida:3OB overestimation of excitation energies in saturated molecules are not nearly as discussed as the self-interaction error of DFT. However, as we will discuss with more detail in a next subsection, this limitation appears to be related to the low quality description of the MOs, due to the use of a minimal basis set, and to the approximations to the charge density in the excited state method. One should be aware that errors related to the non-completeness of the basis set are not exclusive to DFTB, since employing a finite basis set is an approximation present in most numerical methods for the electronic structure of molecules, which of course includes DFT.

Although TD-DFTB appears to be inheriting limitations from its parent formalism, TD-DFT, there are obvious differences. Comparatively, TD-DFT:PBE0/def2SVP can be considered to be more precise for all chemical groups, since the vast majority of its predictions for $E_1$ have an error within $\pm 1~eV$. In contrast, the $E_1$ error dispersion for DFTB-Casida:3OB predictions is highly dependent on chemical identity: it is considerably large for saturated molecules, with $+1~eV < \Delta_{CC2} E_1 < +6~eV$, while for unsaturated molecules it is more moderate, with $\Delta_{CC2} E_1$ ranging between $-2~eV$ and $+1~eV$. Overall, the TD-DFT:PBE0/def2SVP method is expected to be more precise —and also computationally more expensive— because it is less approximated than TD-DFTB: (i) it is an \textit{ab initio} method (i.e., parameter-free), where the expression for the total energy is as exact as the approximate XC-functional being chosen; (ii) it computes the actual XC-functional, which is not necessarily purely local, e.g. PBE0 is a hybrid XC-functional with 25\% exact-HF exchange, and therefore performs significantly better than (semi-)local functionals for valence and CT excitations \cite{Maier2016}; (iii) it employs the def2SVP basis set that, although it is considered to be small for DFT, it is certainly not minimal and even includes some polarization.

Within the unsaturated molecules group, it can be seen that, for molecules characterised by an isolated double or triple bond, DFTB-Casida:3OB populates more than TD-DFT the line at $\Delta_{CC2} E_1 = 0$, which corresponds to the predictions of $E_1$ exactly matching those of CC2. In other words, although the multiple approximations of DFTB to DFT make $E_1$ predictions less accurate, they also led to an increase in cases where errors might be being compensated for to produce the same predictions as methods with a higher level of theory.

While TD-DFTB and TD-DFT are strongly related in formalism, DFTB-ppRPA has a very different conceptual origin. DFTB-ppRPA is expected to perform better than TD-DFT(B) for $\pi$-type excitations, and specifically for excitations with charge transfer (CT) nature, because it manages to reproduce the correct asymptotic $-1/R$ trend for the electron-hole interaction in the long range, even within the monopole approximation \cite{Yang2017}. Because of this, in panel~B of Fig.~\ref{Figure1} it can be seen that DFTB-ppRPA:3OB is not underestimating $E_1$ for all of the unsaturated molecules, as Casida-DFTB:3OB and TD-DFT:PBE0/def2SVP do. In fact, DFTB-ppRPA:3OB overestimates most of the non-conjugated unsaturated molecules and, for $\pi$-conjugated molecules, this method results in a $E_1$ error distribution that is centered at $\Delta_{CC2} E_1 = 0$, although with considerable dispersion. Regarding the saturated molecules, we observe an $E_1$ error distribution that is very similar to that of DFTB-Casida:3OB, suggesting that this limitation originates already in the base formalism, SCC-DFTB.

In the ESI we extended the data analysis of this initial diagnostic to include another layer of chemical detail for the non-conjugated unsaturated molecules; there, we show that DFTB-ppRPA performs best for non-conjugated alkenes, while DFTB-Casida would be the recommended method for the other non-conjugated molecules with carbonyl, cyano and alkyne groups.

Based on the discussion so far, neither of the two approaches based on the standard SCC-DFTB formalism would be acceptable for predicting $E_1$ for saturated molecules, and therefore we should avoid them. With respect to the unsaturated molecules, DFTB-ppRPA generated decent $E_1$ predictions for alkenes and molecules with $\pi$-conjugation, while DFTB-Casida:3OB did so for non-conjugated ones. However, there are extensions to the standard SCC-DFTB formalism, which are already implemented in \texttt{DFTB+}, that are worth testing if we want to improve our initial estimates of $E_1$. This is what we did in hope of resolving the limitations behind the broad overestimation of excitation energies in saturated molecules and behind the underestimation of $\pi$-type excitations, particularly for $\pi$-conjugated molecules. We decided to focus on improving DFTB-Casida from now on, as it is more general than the current implementation of DFTB-ppRPA, which necessarily involves the highest occupied molecular orbital (HOMO) in the prediction of electronic excitations.

\subsection{Dealing with the limitations}

\subsubsection*{Underestimation of $E_1$ for $\pi$-conjugated molecules}

This work is not the first reported encounter of the SIE in DFTB. This DFT-inherited error has already been detected before in the standard formulation of SCC-DFTB \cite{Hourahine2007, Lundberg2011} and was traced back to the approximations made to the charge-fluctuation interaction term, where the XC-component of the charge-charge interaction term is reduced to a single number per element, i.e. the on-site Hubbard-like parameters $U_A$. The highly local and variational nature of the SCC term results in interactions not converging at large distances, in stark contrast to HF or non-SCC DFTB. Because of the SIE, the standard formulation of SCC-DFTB ends up underestimating the energy of delocalised solutions, almost as much as any other GGA functional for DFT (e.g. PBE) \cite{Lundberg2011}. As attempts to overcome this limitation, diverse extensions to the standard SCC-DFTB were proposed, e.g.: LDA+U-DFTB \cite{Hourahine2007}, coarse-grained DFTB \cite{Kubar2010} and configuration interaction (CI) within SCC-DFTB \cite{Rapacioli2011}. Arguably, the most popular and widely accepted strategy to correct the SIE is to include long-range corrections (LC) into SCC-DFTB.

For the benchmark of this correction, we computed $E_1$ with DFTB-Casida+LC and the OB2 parameter set, which is currently the only SK-set available in \texttt{DFTB+} that is compatible with DFTB+LC. In panel~C of Fig.~\ref{Figure1} we can observe that DFTB+LC-Casida:OB2 increases the value of all the initial estimates of $E_1$, resulting in the shift of all corresponding error distributions to more positive values. This results in an improvement for $\pi$-conjugated molecules as their new $E_1$ error distribution is now centered closer to $\Delta_{CC2} E_1 = 0$ and mostly contained within $\pm 1 eV$. It is interesting to note that including long-range corrections into DFTB-Casida (DFTB+LC-Casida:OB2) calculations achieves similar accuracy to DFTB-ppRPA:3OB, but with less dispersion, as can be seen in Fig.~S4 of the ESI. This improved accuracy of DFTB+LC computations arise from the newly gained ability to reproduce the correct trend for interactions over long distances, as DFTB-ppRPA does. This correction being included already from the parameterisation process could be the reason why DFTB+LC-Casida:OB2 is more precise than DFTB-ppRPA:3OB, despite their very similar accuracies. Given that this correction is included via the XC functional, it would be expected for the pre-computed integrals and repulsion splines of DFTB+LC to encode additional XC information. In contrast, DFTB-ppRPA:3OB is based on parameters generated with the non-corrected PBE functional, which is short-ranged and purely local.

The LC shifted the $E_1$ predictions to higher energies for all the molecules in the dataset. Although it resulted in a marked improvement for the initially underestimated $E_1$ predictions (i.e., for the $\pi$-conjugated molecules), it also worsened the estimates for molecules that were already overestimated. Therefore, as can be seen in the projected histograms of Fig.~\ref{Figure1}~C, this corrected method cannot be recommended for non-conjugated unsaturated or saturated molecules.

In the ESI, we provide an extension of the data analysis with an extra layer of chemical detail; we make a distinction between unsaturated molecules with $\pi$-conjugated systems involving 3~atoms and molecules with more extended $\pi$-systems. We found that DFTB+LC-Casida:OB2 performs particularly well for the subset of unsaturated molecules with $\pi$-conjugated systems involving only 3~atoms, of which we can highlight the ester, carboxylic acid and amide functional groups.

Based on the results of this benchmark, we recommend using DFTB+LC-Casida:OB2 for $\pi$-conjugated molecules. This recommendation should become more relevant when studying electronic excitations between different molecular fragments (e.g. solvated chromophores \cite{Isborn2013}), in conducting systems or molecules with very large $\pi$-conjugated systems (e.g. polyacenes \cite{Kranz2017}), and in any other study case where the SIE is expected to be significant.

\subsubsection*{Overestimation of $E_1$ for saturated molecules}

We could identify two main sources for the error that is causing the overestimation of excitation energies in saturated molecules. The first source of error is truncating the multipole expansion of the charge density for the computation of the coupling matrix elements in Casida TD-DFTB. The second source of error is employing a minimal basis set, which results in a low quality description of the GS-MOs. At first glance, what these two approximations have in common is contributing to an over-simplification in the description of the ground state density, either in the ground or excited state formulations.

In the ESI, we listed the three approximations that TD-DFTB makes to LR-TD-DFT. The first two of these approximations are required to translate the Casida's eigenvalue problem to the DFTB framework, one of which is the Mulliken monopole approximation to the transition charge density. The third approximation of TD-DFTB is introduced to minimise the effort in the computation of the coupling matrix elements, by neglecting the charge fluctuations contributions. The latter approximation was considered to be safe by Niehaus \textit{et~al.} \cite{Niehaus2009}, having found that the inclusion of $\Delta q_A$-dependent Hubbard-like $U_{A}$ parameters had a negligible effect on the corrected excitation energies. However, the same cannot be said for the Mulliken monopole approximation, since it becomes troublesome when dealing with highly localised excitations (i.e., with near full orbital overlap), for which it predicts heavily underestimated transition charges. Because same-center atomic orbitals do not contribute to the atomic transition charges within the monopole approximation, in the cases of full orbital overlap these charges vanish and, consequently, so does the coupling matrix; therefore, in those cases TD-DFTB ends up predicting excitation energies that are exactly equal to the KS orbital energy difference.
\noclub[3]

All things considered, to improve the prediction of excitation energies for saturated molecules, we could attempt (i) to correct the monopole approximation in the ES-method or (ii) improve the ground state estimation of the KS energy gap. The second strategy takes into account systems for which $E_1$ can naturally be expected to be very close to the corresponding orbital energy difference.

On the one hand, the first strategy can be addressed with the inclusion of on-site corrections, which were already implemented in \texttt{DFTB+} by Domínguez \textit{et~al.} On the other hand, to improve the KS energy gap, efforts should be directed at refining the energy estimates for the ground state MOs. This can be done by increasing the size of the basis set, which we essentially consider as a stabilisation of the virtual MOs.

Because the tight-binding approach assumes tightly bound electrons, it is justified to employ a minimal basis set to expand the molecular Hamiltonian eigenstates. This approximation is not often a problem for DFTB with a decent parameterization, as it mostly does very well in terms of accuracy and performance for most of its applications, which frequently concern ground state properties. However, when dealing with the energetics of reactions or excitations, the limitations imposed by the constrained description of the orbital solutions becomes more noticeable. For example, Niehaus \cite{Niehaus2009} detected that, because of the minimal basis set description, KS energy differences were larger for DFTB compared to DFT. Furthermore, Gaus \textit{et~al.} \cite{Gaus2013} recognised that the minimal basis set represented a crucial limit to DFTB's accuracy regarding the prediction of reaction energies. We believe that the incompleteness of the basis set predominantly impacts on the overestimation of the virtual state energies, as the unoccupied MOs would not be sufficiently well described. Because of the Rayleigh theorem for eigenvalues, adding more basis functions necessarily lowers the eigenvalues of the molecular Hamiltonian and, thanks to the variational principle, expanding the basis set is not expected to significantly alter the energies of the occupied MOs. However, increasing the number of basis functions has the potential to significantly stabilise the virtual MOs, as they are always fully determined in shape and energy by the requirement of orthogonality with respect to the variationally-determined occupied MOs \cite{Cook1996}, which are better described with larger basis sets.

The easiest way we thought of extending the minimal basis set, only for performing excited state computations with fixed nuclei, was to add pseudo polarization orbitals only to H atoms. The strategy we provide (see the ESI) for minimally polarizing the minimal basis set (i.e., via the custom SK-parameter set, 3OB(H*)), may not be sophisticated enough, but we believe that it serves as a proof-of-concept approach to argue in favour of employing polarized basis sets when predicting $\sigma$-type excitation energies for saturated systems.

\begin{figure*}
\includegraphics[scale=0.46]{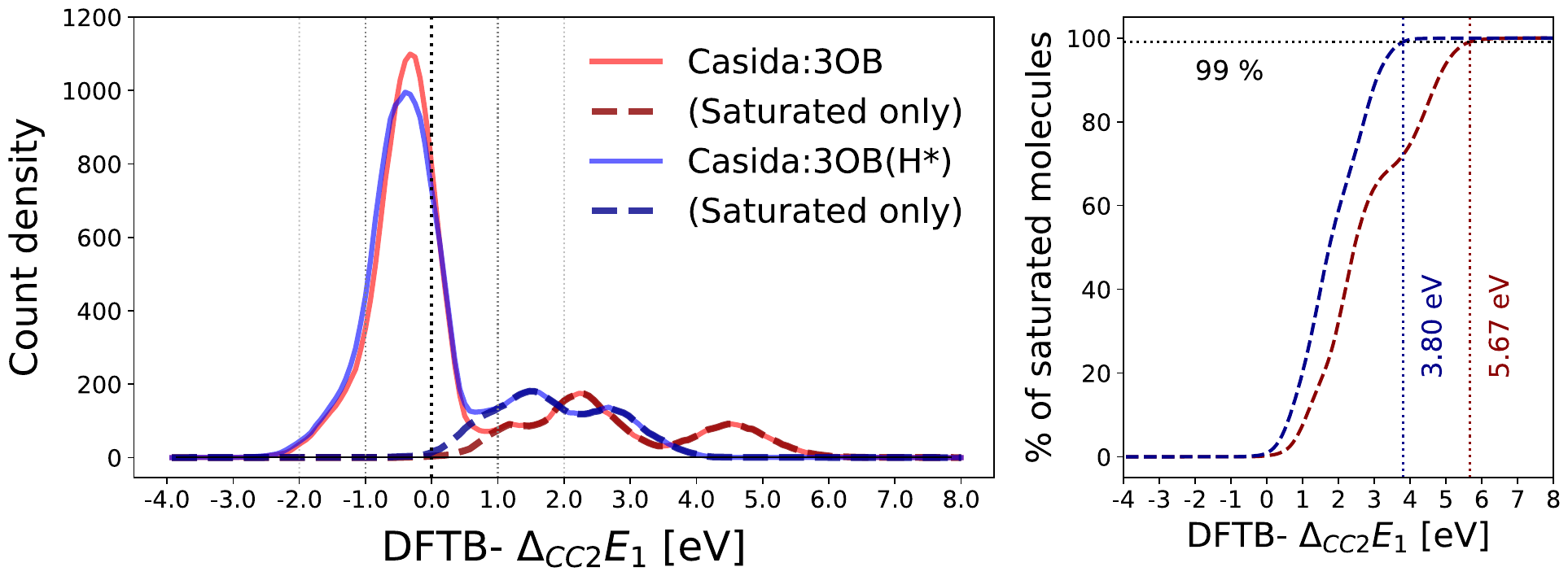}
\caption{\label{Figure2} \textit{Left}: Overlapped $E_1$ error distributions corresponding to all the molecules in the dataset, computed using DFTB-Casida with the original parameter set 3OB (red solid line) and the custom minimally polarized set 3OB(H*) (blue solid line). Histograms of $\Delta_{CC2} E_1$ that correspond only to the subset of saturated molecules are being highlighted as dashed lines in a darker colour. \textit{Right}: Cumulative percentage distribution of saturated molecules with $E_1$ prediction errors bellow $\Delta_{CC2} E_1$; for each method, we highlight the value of $\Delta_{CC2} E_1$ below which 99\% of the saturated molecules are contained.}
\end{figure*}

We benchmarked these two strategies by calculating $E_1$ with the on-site corrected DFTB-Casida+OC:3OB and the minimally polarized DFTB-Casida:3OB(H*). The complete set of results for the on-site corrected ES-method is being displayed in Fig.~S6, in the ESI. Results showing the impact on $E_1$ predictions of partially polarizing the minimal basis set, for all the molecules in the dataset, are given in panel~D of Fig.~\ref{Figure1} and in Fig.~\ref{Figure2}.

The extension of the minimal basis set, to include polarization for H atoms only, would be partially correcting the electron density oversimplification error. As a consequence, DFTB-Casida:3OB(H*) seems to have considerably improved the predictions of $E_1$ for the entirety of the saturated molecules, as the minimal polarization of the basis set resulted in a significant contraction of the $E_1$ error distribution (see Fig.~\ref{Figure2}). Consistently, for the unsaturated molecules, $\pi$-conjugated or not, polarizing the H atoms had an almost negligible effect on their $E_1$ predictions.

Since these two strategies were originally motivated on improving $E_1$ predictions for the saturated molecules, in the ESI we decided to compare this subset of results using an additional layer of chemical detail. From this extended discussion, we found that (i) the inclusion of on-site corrections resulted in acceptable $E_1$ predictions for the subfamily of molecules with an epoxy functional group, and that (ii) partially polarizing the minimal basis set had a larger corrective effect on the subfamilies with more s-character (i.e. ethers, alcohols and saturated hydrocarbons).

However, based on our results we cannot recommend any of the here benchmarked methods to predict $E_1$ for non-epoxy saturated molecules. The custom parameter set 3OB(H*) was only employed to highlight the favourable impact that including polarization orbitals has on $E_1$ predictions, as it improves the GS estimates of the KS energy gaps. At most, what we proposed should serve to assist the developer community of \texttt{DFTB+} in setting the direction for improvement towards a more comprehensive solution. If the community reaffirms its interest in studying the excited state of saturated systems, we recommend as an extra step in \texttt{DFTB+}'s development roadmap the inclusion of extra polarization orbitals for each of the elements in 3OB, what may require a re-parameterization of the two-center repulsion splines. The natural following step would be to extend the base formalism to allow polarization via the multipole expansion of the charge density $\delta \rho$, beyond the zeroth order.

Essentially, we are proposing that future implementations of \texttt{DFTB+} regarding this topic should be based on the Gaussian polarizable-ion tight biding (GTB) method, developed by Horsfield, Boleininger \textit{et~al.} \cite{Boleininger2016}. GTB performs on-the-fly computations of the multipole moments, up to quadrupole order, from a Gaussian basis set with the potential to include polarization orbitals. Currently, an implementation of this method exists within another suite of programs for computational chemistry, named Plato \cite{Horsfield1997, Horsfield2000, Kenny2009}. For a reduced set of small hydrocarbon molecules, GTB predicted mean polarizability volumes ($\alpha_{m}$) in excellent agreement with experimental determinations, similarly to DFT-PBE but at a fraction of the computational cost. We focused on a subset of these results, comprising H$_2$ and the first four elements of the homologous series of straight-chain alkanes (shown in Table~II of our ESI). For these saturated molecules, in order to significantly counteract the sharp underestimation of $\alpha_{m}$, leading to the discussed overestimation of the $\sigma$-type $E_1$, it is necessary to append polarization orbitals to the minimal basis set, since the multipole expansion of the charge density alone is apparently not sufficient. For GTB with a second-order expansion of the charge density, the mean relative absolute percentage error of $\alpha_{m}$ goes from approximately 68\% to near 8\%, just with the addition of polarization orbitals for each element (i.e., p orbitals for H and d orbitals for C); for comparison, this error was close to 5\% for DFT:PBE/cc-pVQZ.

Lastly, it is worth mentioning another alternative to improve $\sigma$-type excitation energies, which we do not cover in this work as it is not implemented into \texttt{DFTB+}. It consists on extending SCC-DFTB to include the chemical-potential equalisation (CPE) approach, in which an additional response density is supplied in order to overcome the limitations of the minimal basis set with respect to response properties \cite{Kaminski2012}.

\section{Summary and Conclusions}

With this work, we support the idea that there are lessons to be learned from the analysis of large amounts of chemical data, if we complement the datasets with relevant molecular descriptors and an adequate theoretical background, even when the datapoints are not as detailed as would be ideal. For example, our data analysis focused only on $E_1$ values \footnote[4]{In the ESI, we also performed a brief analysis on the oscillator strengths associated with $E_1$ (i.e., $f_1$) and found results that complement the conclusions of this work.} from a ML-dataset and did not take into account any other information related to the electronic transitions, such as the symmetry of the excited states, as this extra level of detail is not always available \footnote[5]{Currently, \texttt{DFTB+} does not support point group symmetry.}. Although this lack of additional information may have acted as a source of a systematic underestimation error for $E_1$, which can alternatively be viewed as a qualitative error in terms of symmetry \footnote[6]{For the cases where DFT- or DFTB-based methods predict an excitation energy that exactly matches the reference value (i.e., of CC2), the predicted symmetry of the corresponding ES may not match the reference, as it is possible for the energy ranking of the states to change with the method.}, we were able to extract robust and valuable insights on the limitations of the benchmarked ES-methods. For this task, taking into account the chemical identity of the molecules and employing a sufficiently large and chemically diverse ML-dataset were key aspects. 

With our chemically-informed analysis of the GDB-8 chemical subspace, we were able to identify clear trends in the prediction error distributions for $E_1$. With this benchmark we learned that, despite being more approximate, DFTB ES-methods suffer from the same limitations as the parent method, DFT, in the prediction of $E_1$, albeit to a greater extent. On the one hand, there is the self-interaction error, which mainly affects $\pi$-type excitations in molecules with $\pi$-conjugated systems or high p-character. On the other hand, there is the error arising from the over-simplification in the description of the ground state density (e.g., due to the use of a non-complete basis set), which has a significant impact on $\sigma$-type excitations, in molecules with predominantly s-character. However, in this work we showed that we can choose between the currently implemented methods and corrections of \texttt{DFTB+} in order to predict $E_1$ with errors mostly within $\pm 1~eV$. Beware that, as this work bases its conclusions on general trends, any system-specific analysis on the correctness of the energy and symmetry ranking of the low-lying ES becomes delegated to the researchers interested in studying such particular systems.

\begin{figure*}
\includegraphics[scale=0.55]{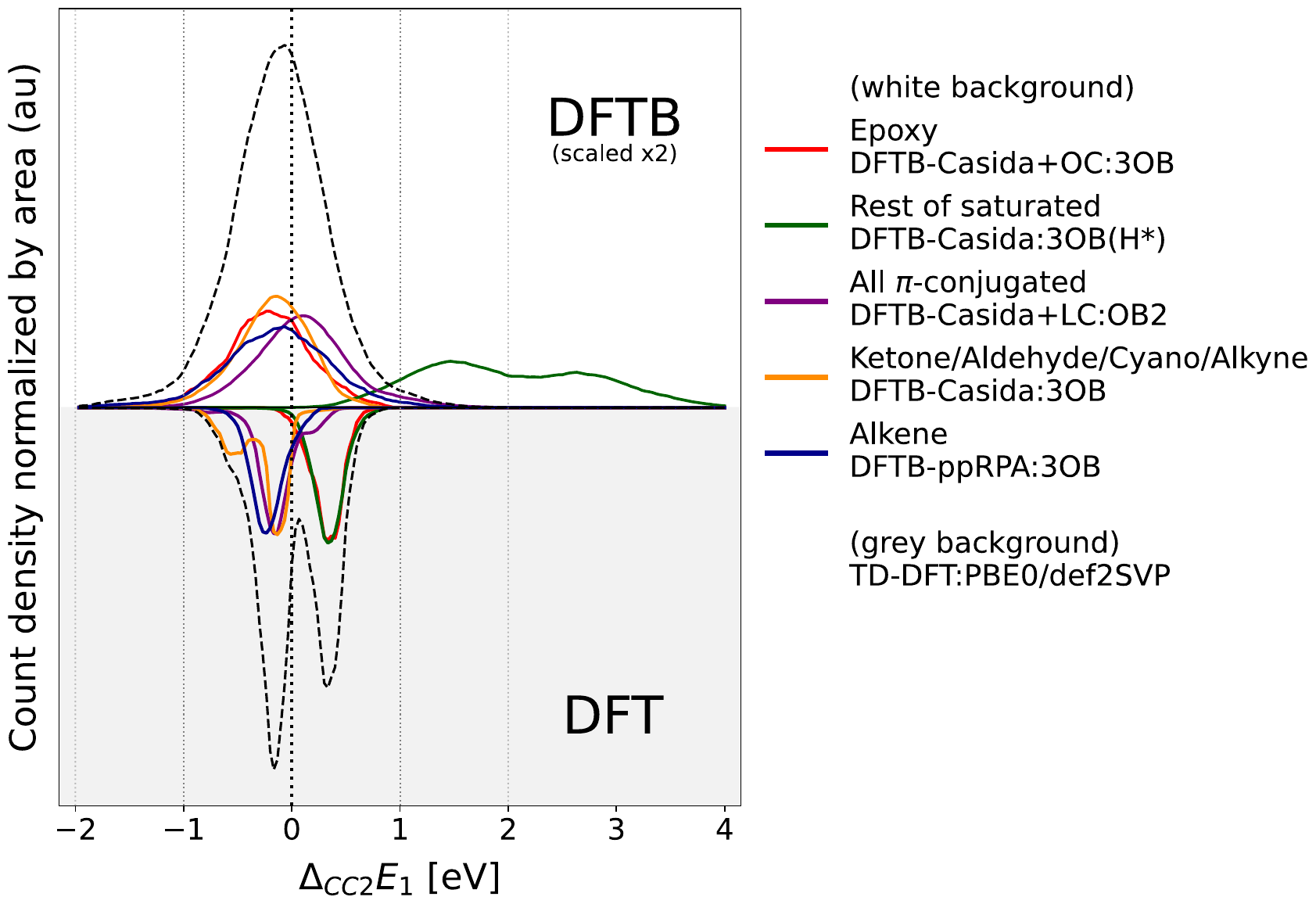}
\caption{\label{Figure3} Comparison of $E_1$ error distributions ($\Delta_{CC2} E_1$) normalised by area, for chemical subfamilies of molecules from the dataset, computed using different DFTB methods (at the top, on a white background) and TD-DFT:PBE0/def2SVP (at the bottom, on a grey background). To compute $E_1$ with \texttt{DFTB+} we followed the rule of thumb we constructed throughout the discussion of results. Accordingly, we employed: DFTB-Casida+OC:3OB for saturated molecules with epoxide groups (red solid line), DFTB-Casida:3OB for the rest of the saturated molecules (green solid line), DFTB-Casida+LC:OB2 for all $\pi$-conjugated unsaturated molecules (purple solid line), DFTB-Casida:3OB for non-conjugated unsaturated molecules with carbonyl (i.e., ketones or aldehydes), cyano or alkyne groups (orange solid line) and DFTB-ppRPA:3OB for unsaturated molecules with isolated alkene groups (blue solid line). The combined error distributions for DFT and DFTB methods are shown (as dashed black lines) superimposed on their respective sides of the plot. The combined error distribution for the DFTB-approximate methods ignores the non-epoxy saturated molecules, as we could not recommend any suitable implementation within \texttt{DFTB+} for this subgroup. For better visualisation, all DFTB distributions were scaled to double. See the first column of Tables~IV~\&~V, in the ESI, for the number of occurrences per chemical family and subgroup.}
\end{figure*}

Our benchmark allowed us to construct a ``rule of thumb'' intended to assist the community of users in selecting the most reliable DFTB approximation for the ES calculations of their molecular systems of interest. According to this rule, visually summarised in Fig.~\ref{Figure3}, when computing $E_1$ with \texttt{DFTB+} we recommend using: (i)~DFTB-Casida+OC:3OB for saturated molecules with epoxide groups, (ii)~DFTB-Casida+LC:OB2 for $\pi$-conjugated unsaturated molecules, (iii)~DFTB-Casida:3OB for non-conjugated unsaturated molecules with carbonyl (i.e., ketones/aldehydes), cyano or alkyne groups, and (iv)~DFTB-ppRPA:3OB for unsaturated molecules with isolated alkene groups. Ignoring non-epoxy saturated molecules, the combined error distribution resulting from applying the proposed rule of thumb is almost as acceptable as that of the computationally more expensive TD-DFT method with a corrected exchange functional (PBE0) and a partially polarized basis set (def2SVP). Surprisingly, Fig.~\ref{Figure3} also shows that, in contrast to TD-DFT, the combined error distribution of the DFTB-approximate methods is nearly centred on $\Delta_{CC2} E_1 = 0$, i.e. the line corresponding to exact matches with CC2 predictions. As previously discussed, we believe that, since DFTB's predictions of $E_1$ appear to be quite sensitive to the identified sources of error, the occurrence of beneficial error compensation increases. Although we cannot recommend any DFTB-approximate method within DFTB+ to predict $E_1$ for most saturated molecules, if the community reaffirms its interest in studying the excited state of these systems, we recommend adding the following steps to the development roadmap of DFTB+: (i) the inclusion of extra polarisation orbitals in 3OB and (ii) the extension of the base formalism to allow for polarisation via the multipole expansion of the charge density, beyond zeroth order. This will require referring to the GTB method by Horsfield, Boleininger \textit{et~al.} \cite{Boleininger2016}.

\section*{Acknowledgments}
The authors acknowledge financial support by the Consejo Nacional de Investigaciones Cíentificas y Técnicas (CONICET) and the Agencia Nacional de Promoción Científica y Tecnológica (Agencia I+D+i), Argentina, through grants PIP 112-2017-0100892CO and PICT-2017-1506. A. I. B. also thanks CONICET for the doctoral fellowship. Both authors would like to extend a special thanks to Dr. Matías Berdakin (CONICET) for the very helpful feedback. We would also like to thank all the anonymous reviewers who contributed to the overall quality of this manuscript by providing constructive comments after a critical reading.

\bigskip
\nocite{*}
\bibliography{main}

\end{document}


\title{Electronic Supplementary Information (ESI) for\\``Data-driven approach for benchmarking DFTB-approximate excited state methods''}

\author{Andrés I. Bertoni}
\author{Cristián G. Sánchez$^{*}$}
\email[Corresponding author.~E-mail:~]{csanchez@mendoza-conicet.gob.ar}
\affiliation{Instituto Interdisciplinario de Ciencias Básicas (ICB-CONICET), Universidad Nacional de Cuyo, Padre Jorge Contreras 1300, Mendoza 5502, Argentina}

\maketitle

\section{Overview of methods}
\subsection{SCC-DFTB approximations to KS-DFT}

We can completely characterise the self-charge-consistent version of DFTB (SCC-DFTB) by listing its approximations to Kohn-Sham's DFT (KS-DFT). This set of additional approximations to the parent framework are key to its superior efficiency.

DFTB is derived from a truncated expansion of the KS-DFT total energy functional. In this work, we made use of SCC-DFTB, which includes terms up to the second order in the expansion of the ground-state electron density~$\rho$, around a reference density~$\rho_{0}$ being perturbed by density fluctuations~$\delta \rho$:
$$ E^{\scalebox{0.6}{\textrm{SCC-DFTB}}} \left[ \rho_{0} + \delta \rho \right] = E^{0} \left[ \rho_{0} \right] + E^{1} \left[ \rho_{0}, \, \delta \rho \right] \\ + E^{2} \left[ \rho_{0}, \, \left( \delta \rho \right)^{2} \right] $$
The truncation that leads to the above expression constitutes the \textit{first approximation} of SCC-DFTB to the exact KS-DFT total energy functional. Since~$\rho_{0}$ is typically constructed as a superposition of neutral atomic densities, $\delta \rho$~would be accounting for the chemical environment of each atom within the molecule. In approximate KS-DFT, the molecular orbitals are expanded in the basis of atom centered basis functions, i.e as a linear combination of atomic orbitals (LCAO):
$$ \psi_{i} \left( \boldsymbol{r} \right) = \sum_{\mu} c_{\mu}^{i} \phi_{\mu} \left( \boldsymbol{r} \right) $$
This projection on atom-centered basis functions converts the time-independent Schrödinger equation into an eigenvalue problem. As a \textit{second approximation}, DFTB's standard formulation employs a minimum set of valence orbital basis functions (i.e. a minimal basis set) to simplify the linear algebra operations. In practice, these basis functions are slightly compressed atomic-like solutions to the KS-DFT equations. It should be cautioned that, as standard SCC-DFTB uses a minimal basis set, Rydberg states are outside the scope of DFTB-approximate ES-methods; this particular type of electronic excitations would require, instead, the use of a very diffuse basis set to be described correctly.

In the expansion of the total energy, the zeroth-order term~$E^{0}[\rho_{0}]$ is the energy associated with the repulsion interaction between the nuclei and between the atomic contributions to the reference density~$\rho_{0}$. The \textit{third approximation} of DFTB is to write~$E^{0}[\rho_{0}]$ as a sum of pair-potentials:
$$ E^{0}[\rho_{0}] \approx \frac{1}{2} \sum_{AB} V_{AB}^{rep} \left( \boldsymbol{R}_{AB} \right) $$
The first-order term~$E^{1} \left[ \rho_{0}, \, \delta \rho \right]$ is the band-structure energy, which involves the computation of matrix elements of the reference Hamiltonian~$H \left[ \rho_{0} \right]$:
$$ E^{1} \left[ \rho_{0}, \, \delta \rho \right] = \sum_{i} f_i \sum_{\mu \nu} c_{\mu}^{i*} \, c_{\nu}^{i} \, H_{\mu \nu}^{0} \quad \mu \in A, \nu \in B $$
where~$f_i$ are the Fermi occupations of the ground-state molecular orbitals within the LCAO \textit{ansatz}. Obtaining these matrix elements require the computation of three-center integrals, because they involve atomic orbitals from two different centers, $\phi_{\mu} \left( \boldsymbol{r} \right)$~and~$\phi_{\nu} \left( \boldsymbol{r} \right)$, and the effective potential at the reference density, $V_{s} \left[ \rho_{0} \right] \left( \boldsymbol{r} \right)$. The \textit{fourth approximation} of DFTB is to neglect the two-center crystal-field contributions and the three-center interactions in the band energy term, transforming the reference Hamiltonian diagonal and non-diagonal elements, $H_{\mu \mu}^{0}$~and~$H_{\mu \nu}^{0}$, into less expensive one- and two-center integrals, respectively.

Additionally, SCC-DFTB makes two approximations to the energy from charge fluctuations, i.e. the second-order term in the total energy expression ($E^{2} \left[ \rho_{0}, \, \left( \delta \rho \right)^{2} \right]$). The \textit{fifth approximation} of SCC-DFTB is the monopole approximation of the ground state (GS) charge density $\delta \rho$. The zeroth-order truncation of the multipole expansion of~$\delta \rho$ neglects the one-center two-electron integrals and introduces the computation of partial atomic charges~$q_{A}$ (and atomic populations~$\Delta q_{A}$) from two-center orbital overlap matrix elements~$S_{\mu \nu}$, under the Mulliken population analysis:
$$ \Delta q_{A} = q_{A} - q_{A}^0 $$
where
$$ q_{A} \equiv \sum_{i} f_i \sum_{\mu \in A} \sum_{\nu} \frac{1}{2} \left( c_{\mu}^{i\,*} \, c_{\nu}^{i} \, S_{\mu \nu} + c_{\nu}^{i\,*} \, c_{\mu}^{i} \, S_{\mu \nu} \right) $$
and~$q_{A}^{0}$ is the number of valence electrons in neutral atom~$A$. Notice that the charge-fluctuation energy term is the one turning DFTB into a SCC method, since the atomic populations themselves depend on the molecular orbital coefficients. Lastly, SCC-DFTB's \textit{sixth approximation} is to enforce the locality of XC contributions, making the charge-fluctuation interaction only electrostatic (i.e. Coulombic) for different centers. The interactions between the atomic populations follows a dependence with the inter-atomic distance given by an analytical density profile function~$\gamma_{AB}$ (e.g. a Gaussian profile) that becomes equal to the Hubbard~$U_{A}$ parameter when~$A=B$, which is an on-site contribution linked to the chemical hardness of the atom:
$$ E^{2} \left[ \rho_{0}, \, \left( \delta \rho \right)^{2} \right] \approx \frac{1}{2} \sum_{AB} \gamma_{AB} \left( R_{AB} \right) \, \Delta q_{A} \, \Delta q_{B} $$
A very strong point towards the superior computational performance of DFTB is the pre-computation of matrix elements and repulsion pair-potential splines. With this strategy, the cost associated with having to compute the approximate integrals is transferred to a one-time parameterization process. The Hamiltonian and overlap matrix elements~$H_{\mu \mu}^{0}$ and~$S_{\mu \nu}$, needed for the first- and second-order energy terms, are highly transferable as they are pre-determined for a reference density. Within the standard SCC-DFTB, these electronic parameters are pre-computed for various interatomic distances using a minimal basis set of pseudo-atomic KS-orbitals, obtained from atomic DFT calculations with a confining potential and a local-XC functional, such as the parameter-free Perdew-Burke-Ernzerhof (PBE) functional of the generalised gradient approximation (GGA) class. In an independent stage of the parameterization process, the zeroth-order repulsive pair-potentials are most often fitted to results from DFT calculations \cite{Porezag1995, Elstner1998, Gaus2013, Vuong2018} for a set of element pairs, but these functions can also be fitted to experimental data (e.g., equilibrium geometries, atomization energies and vibrational frequencies) \cite{Gaus2009}. When DFT is used to fit the repulsive functions, calculations are expected to be performed close to the basis set limit and with high-quality functionals. These repulsion functions are the parameters that make DFTB a semi-empirical method and, in standard SCC-DFTB, are expected to encode all the chemically relevant non-local nature of the electron-electron interaction. The Hubbard~$U_{A}$ parameters are also obtained from DFT pre-computations. All these numbers are generated via the Slater-Koster (SK) technique \cite{Slater1954} and stored as SK-files for different pairs of chemical elements; this set of files is what we call a parameter set in DFTB.

It is also possible to improve the first approximation and extend the expansion of the total energy to include higher-order energy terms. \texttt{DFTB+} can perform computations up to the third-order in energy. However, this extra energy term only becomes important when bonding results in large atomic charge fluctuations, i.e. for local densities deviating significantly from the reference one.

\subsection{TD-DFTB from Casida's LR-TD-DFT}

The TD-DFTB method for computing excitation energies is based on Casida's LR-approach \cite{Casida1995} to TD-DFT, the time-dependent extension of the Hohenberg-Kohn theorems. In the dynamic LR treatment of the GS electron density being perturbed by an external electric potential, Casida derived a pseudo-eigenvalue equation in the space of single-orbital transitions from $i,j$-occupied to $a,b$-virtual molecular orbitals:
$$ \sum_{jb} \left[ \omega_{ia}^{2} \delta_{ij} \delta_{ab} + 4 \sqrt{\omega_{ia}} K_{ijab} \sqrt{\omega_{jb}} \right] F_{I}^{jb} = \omega_{I}^{2} F_{I}^{ia} $$
where the KS orbital energy gaps, $\omega_{ia}$, and the coupling matrix elements, $K_{ijab}$, are constructed from the ground state. This problem can be solved for a number of $I$ electronic transitions, in order to determine their excitation energies $\omega_{I}$ and transition contributions $F_{I}^{ia}$, which are needed to compute the corresponding transition dipole moments and approximate excited state wavefunctions. It can be noticed that $\omega_{I}$ results from the correction made by the coupling matrix, which accounts for the electron-hole interaction, to the initial excitation energy estimate given by the KS energy difference. Furthermore, $\boldsymbol{F}_{I}$ being a vector indicates that electronic transitions can display multi-orbital character.

TD-DFTB is a translation of the Casida's eigenvalue problem to the DFTB framework. It consists on extending SCC-DFTB approximations to significantly simplify the computation of the coupling matrix elements $K_{ijab}$. The one-center (or on-site) exchange-like integrals are neglected and the remaining expensive two-center two-electron integrals are converted into simple sums for atom pairs. The following is the resulting approximate expression valid for singlet-singlet transitions:
$$ K_{ijab} \approx \sum_{AB} \gamma_{AB} \left( R_{AB} \right) \; q_{A}^{ia} \; q_{B}^{jb} $$
where 
$$ q_{A}^{ia} \equiv \sum_{\mu \in A} \sum_{\nu} \frac{1}{2} \left( c_{\mu}^{i\,*} \, c_{\nu}^{a} \, S_{\mu \nu} + c_{\nu}^{i\,*} \, c_{\mu}^{a} \, S_{\mu \nu} \right) $$
are transition charges introduced by the Mulliken monopole approximation. The XC term in the exact expression for the coupling matrix elements $K_{iajb}$ depends on the full ground-state electron density $\rho$. That means that the XC contribution not only depends on the reference density, as in the second-order energy term, but also on the density fluctuations from the $\rho_{0}$ GS-reference. This would require promoting the Hubbard $U_A$ parameters into functions of the site's atomic population, i.e. $U_A \left( \Delta q_{A} \right)$. However, a third approximation is made to TD-DFTB in order to neglect these charge fluctuations when computing the coupling matrix elements, allowing reuse of the approximate profiles $\gamma_{AB}$ from the initial GS computations.

It is worth pointing out that, as Niehaus has previously shown \cite{Niehaus2009}, the results from the DFTB-approximate Casida eigenvalue problem are completely equivalent to those extracted from the real-time TD-DFTB approach, for which an implementation in \texttt{DFTB+} was reported by Bonafé \textit{et~al.} \cite{Bonafe2020}. The latter approach may pose a challenge in the deconvolution of excitation peaks for spectroscopically complex systems, but it has some advantages over TD-DFTB-Casida: (i) it does not require truncating the number of excitations to be computed, (ii) it can result in a superior computational performance for very large systems, and (iii) it can be extended for the calculation of transient absorption spectra (TAS) simulations. For small systems, like the ones we are employing for this benchmark, TD-DFTB-Casida was the most convenient choice in terms of performance and output parsing.

\subsection{DFTB-approximate ppRPA}

Borrowed from nuclear physics, ppRPA is an eigenvalue problem describing 2-electron addition and removal processes. It can be used to predict the excitation energies of N-electron systems via the treatment of two-electron additions in corresponding two-electron deficient (N-2) systems:
$$ \omega_{0 \rightarrow n} = \omega_{n}^{+2} - \omega_{0}^{+2} $$
where $\omega_{0}^{+2e}$ and $\omega_{n}^{+2e}$ are the eigenvalues for the ground state and the $n$-excited state, respectively. The above expression can be better understood with the help of the diagram in Fig.~S1 of this ESI. Once again, the translation into the DFTB framework consists on the approximation of bottleneck integrals. Originally, the matrix elements to be calculated before solving the eigenvalue problem each involve two integrals of two electrons. In DFTB-approximate ppRPA, the Mulliken monopole approximation reduces these expensive integrals to summations of simpler terms, which share strong similarities with the expression for the coupling matrix elements in TD-DFTB.

\section{Extensions to the standard \\ SCC-DFTB method}
\subsection{Including long-range corrections in TD-DFTB}

To better account for non-local contributions and reproduce the correct -1/R asymptotic trend for Coulombic interactions in the long distance, parameter sets for DFTB can be constructed with range-separated or long-range corrected (LC) XC functionals. Within this scheme, the two-electron interaction is split into short-range and long-range components, with the splitting being modulated by the range-separation parameter~$\omega$:
$$ \frac{1}{r} = \frac{1-e^{-\omega \, r}}{r} + \frac{e^{-\omega \, r}}{r} $$
The implementation of DFTB+LC \cite{Niehaus2001, Niehaus2012, Lutsker2015} into \texttt{DFTB+} is quite recent and therefore there is only one openly available parameter set prepared to include these corrections into DFTB: OB2(-1-1) \cite{Vuong2018}. This SK-set employs the range-separation parameter~$\omega = 0.3 \, a_0^{-1}$ and manages to reproduces GS geometries and vibrations of CHON organic molecules with a similar quality as DFTB3:3OB \cite{Gaus2013}. Thanks to the \texttt{DFTB+} community being actively working to extend existing parameter sets, there is reported a very recent re-parameterisation of OB2 to include sulphur heteroatoms in organic molecules \cite{Varella2021}. Yet, we could not find any other reported extension of OB2 and therefore we decided to ignore the fluorinated organic molecules for our DFTB+LC calculations, a subset that only represents about~$1.4 \, \% $ of all the molecules in the dataset.

\subsection{Including on-site corrections in TD-DFTB}

The standard SCC-DFTB formalism can be extended to partially correct the monopole approximation of the transition charge density in Casida TD-DFTB, in order to no longer neglect the on-site integrals of the exchange type. This implementation is known in \texttt{DFTB+} as on-site corrections (OC) \cite{Dominguez2013}. This correction requires providing additional on-site constants that depend only on the XC-kernel, which we extracted from the Appendix~J of the \texttt{DFTB+} manual \cite{DFTBManual}. Once again, we have ignored fluorinated molecules for the DFTB+OC computations, as there are no available pre-computed on-site constants for fluorine atoms.

\subsection{Partially polarizing the minimal basis set}

The standard formulation of DFTB is characterised by the use of a minimal basis set. However, we could instead employ an extended, yet limited, basis set for the pre-computation of the electronic integrals in the parameter set SK-files. With the \texttt{SkProgs} package \cite{SkProgs} for \texttt{DFTB+}, we constructed a proof-of-concept, custom SK-parameter set that works with a minimally polarized minimal basis set. Since the electronic part of this custom SK-parameter set was intended to emulate an extension of the 3OB parameter set \cite{Gaus2013} to include minimal polarization on H atoms only, we decided to name this set ``3OB(H*)''.

To achieve this minimally polarized set, we added empty $2p$~orbitals to provide extra angular degrees of freedom to the valence electrons on the hydrogen centers and break the original radial symmetry of the $1s$~orbital. The radial dependence of the extra polarization orbitals was scaled to bring it closer to that of the $1s$~valence orbital, but without compromising its original profile. We achieved the custom radial probability densities (see Fig.~S2 of this ESI) by limiting the radial wave function (see equation 7.2 in the \texttt{DFTB+} manual \cite{DFTBManual}) to one variational coefficient $c$ and one exponent $\alpha = \sqrt{2}$:
$$ R_{p}(r) = c \; r \; e^{-\sqrt{2} \, r} $$

As we only intended to perform single-point computations (i.e., with fixed nuclei), it was not necessary to re-parameterise the pair-repulsion splines, which we kept from the original 3OB SK-files.

\section{Further discussion of results}

\subsection{Non-conjugated unsaturated molecules}
In order to construct a better ``rule of thumb'' for the DFTB-approximate ES-methods, we included another layer of chemical detail for the non-conjugated unsaturated molecules. In Fig.~S3 of this ESI, it can be seen that DFTB-Casida:3OB performed best for non-conjugated molecules with carbonyl (i.e., ketones and aldehydes), cyano and alkyne groups, for which their $E_1$ error distributions are centered near $\Delta_{CC2} E_1 = 0$ and mostly contained within $\pm 1~eV$. However, DFTB-Casida:3OB may not be the preferred choice for non-conjugated alkenes, as its estimates of $E_1$ are affected by a systematic underestimation and a significant error dispersion. We suspect that this systematic underestimation of $E_1$ for alkenes is also related to the self-interaction error.

Not all systems are equally affected by the SIE. We expect the SIE-induced underestimation of the delocalised solutions to be more pronounced in molecules with naturally delocalised molecular orbitals of large spatial extent (e.g., $\pi$-conjugated systems). If the delocalised orbitals do participate in low-lying electronic transitions (e.g., frontier orbitals such as HOMO or LUMO), then a systematic underestimation of the corresponding excitation energies is also to be expected. Molecules of the chemical subgroup of alkenes are characterised by the presence of an isolated carbon-carbon double bond that is not part of a conjugated $\pi$-system and, therefore, is rather spatially confined. In the case of alkenes we would have expected a fairly small underestimation of the first excitation energy. However, alkenes appear to be strongly affected by a systematic underestimation of $E_1$.

For systems affected by the SIE, the DFTB-ppRPA method is expected to achieve better results and, as it can be seen in Fig.~S1 of this ESI, this is indeed what is observed for alkenes. It remains to be asked what makes alkenes more susceptible to this error than other compounds with isolated double or triple bonds. We speculate that alkenes may suffer from the SIE-induced artificial stabilisation of delocalised solutions to a greater extent, and thus have underestimated their $E_1$ predictions, as they are more easily polarisable than the other subfamilies of non-conjugated unsaturated molecules, with permanent dipole moments (e.g., carbonyl and cyano groups) or with higher order bonds (i.e., alkynes). In Table~III of this ESI, it can be seen that the measured polarisabilities \cite{Gussoni1998, NIST, Zevatskii2006} of linear alkenes are indeed higher than those of other linear non-conjugated unsaturated molecules of the same length (i.e., with the same number of non-H atoms). On a related side note, the SIE was found to decrease with increasing disparity of electron affinities between the electron and hole regions \cite{Lundberg2011}, which may reinforce the reason why the SIE is lower for carbonyl and cyano groups, where excitations between frontier MOs are expected to involve a non-bonding molecular orbital ($n$) highly localised on the heteroatom (O or N, respectively) and an anti-bonding $\pi$ orbital ($\pi^{*}$) delocalised over the double bond.

\subsection{Unsaturated $\pi$-conjugated molecules}
In Fig~S5 of this ESI, we show that DFTB+LC-Casida:OB2 performs particularly well for the subset of unsaturated molecules with $\pi$-conjugated systems involving only 3~atoms. The $E_1$ error distributions for these chemical subgroups can be seen to be contained between $\Delta_{CC2} E_1 \pm 1~eV$. The most accurate $E_1$ predictions appear to have been achieved for the subgroup of carboximidates; however, we can also highlight the ester, carboxylic acid and amide functional groups, which are known for their chemical importance and ubiquity.

\subsection{Saturated molecules}
Since the inclusion of extensions to the standard SCC-DFTB method was motivated in part on improving the predictions of $E_1$ for the saturated molecules, we decided to compare this subset of results using an extra layer of chemical detail.

The inclusion of on-site corrections resulted in a red-shift of $E_1$ for all saturated molecules containing at least one oxygen or nitrogen heteroatom; the average shift observed for epoxides was of about -1.2~eV, and of approximately -0.9~eV for aziridines, while hydrocarbons (saturated molecules of C and H only) were almost unaffected by this correction. Thanks to the on-site corrections, DFTB-Casida+OC:3OB was able to provide acceptable $E_1$ predictions for the subfamily of molecules with an epoxy functional group.

As seen in Fig.~S7 of this ESI, both the on-site corrected DFTB-Casida+OC:3OB and the minimally polarized DFTB-Casida:3OB(H*) performed best for the subfamilies of saturated molecules containing highly strained three-membered rings, such as epoxide, aziridine and cyclopropyl groups.

These ring structures have markedly acute bond angles and therefore greater p-character than the non-strained saturated molecules. Together with the large orbital overlap that exists at the center of these small rings, $\sigma$-electrons end up delocalising with a stabilising effect on the anti-bonding virtual MOs, in what is known as hyperconjugation \cite{Inagaki1994}. The electron density to be delocalised on the rings can also come from geminal $n$-orbitals (due to an interaction known as negative hyperconjugation) and from electropositive substituents \cite{Wu2013}. The highly strained saturated molecules are a special case. On the one hand, being saturated molecules, their electronic excitations are expected to involve $\sigma$-type occupied molecular orbitals and, therefore, may be in need of a better description of the electron density to avoid overestimating their energies. On the other hand, their higher p-character would give rise to delocalised MOs, particularly virtual states near the frontier, which can potentially be underestimated in energy by the SIE. Therefore, for these subgroups of molecules we would expect cases that give rise to beneficial error compensation.

While it would certainly be interesting to further investigate our hypothesis on error compensation, it is also beyond the scope of this study. At this stage, we limit ourselves to updating our rule of thumb to include the recommendation to use DFTB-Casida+OC:3OB when predicting $E_1$ for the saturated molecules of the epoxy family.

We also calculated the average shifts observed in $\Delta_{CC2} E_1$ after the inclusion of partial polarisation (H*) and of long-range corrections (LC), for the sets of three-membered ring molecules; for each of the chemical groups discussed we have observed smooth monomodal distributions with a full width at half maximum (FWHM) between 0.5~eV and 1~eV (not shown).

On a hand, the addition of partial polarisation red-shifted $E_1$ in all cases, suggesting that the electron density oversimplification error was present. We observed the largest mean shift for cyclopropanes (-0.66~eV), followed by aziridines (-0.53~eV) and epoxides (-0.40~eV); we correlated this with the number of geminal lone pairs in each chemical group (epoxides have two because of oxygen, aziridines have one from nitrogen, and the cyclopropyl group has none). We would expect molecules with fewer valence p-electrons to rely more on the quality of the electron density of $\sigma$ bonds for their $E_1$ predictions.

On another hand, the inclusion of long-range corrections resulted in a detrimental blue-shift for all $E_1$ predictions, which may signal that the SIE was present. Again, we observed the largest average shift for cyclopropanes (+2.15~eV), but this time it was followed by epoxides (+1.85~eV) and then aziridines (+1.48~eV). We previously conjectured that systems with more delocalised solutions and higher polarisabilities were expected to be more affected by the SIE. To compare the polarisabilities of these three chemical subgroups we can refer to Table~III, available in this ESI. In this table, we can see that there is a trend in polarisabilities based on elemental composition: hydrocarbons are more polarisable than molecules with oxygen, and both are more polarisable than nitrogen-containing molecules. Although the discussed compounds are cyclic, and solely comprised of single bonds, we can think of three-membered rings as being more similar to their very short (2 and 3 non-H atoms) linear analogues.

\vspace{2 em}
\subsection{Regarding oscillator strengths}
In Fig.~S8 of this ESI the following can be noticed: (i) for saturated molecules, $f_1$ prediction errors are present almost exclusively for the DFTB-approximate methods (see the datapoints populating the horizontal line at $y=0$); (ii) for $\pi$-conjugated molecules with more than 3~conjugated atoms, the underlying limitation is shared between the compared methods, in a considerable amount of cases (see the datapoints along the diagonal line at $y=x$); (iii) for $\pi$-conjugated molecules with 3~conjugated atoms and for the unsaturated non-conjugated molecules, $f_1$ prediction errors are present but are rather mild (see the datapoints clustered near the origin).

This same information can be extracted from the information compiled in Tables~IV~\&~V of this ESI, where we also provide a second layer of chemical detail (i.e., a breakdown of results into the different subgroups within each main chemical family). One additional observation that we can make to the values in the aforementioned tables is that, for molecules with $\pi$-conjugation, there is a tendency for the accuracy of $f_1$ predictions to worsen with increasing number of atoms involved in the $\pi$-conjugated system.

\bigskip
\nocite{*}
\bibliography{supp}

\setlength{\tabcolsep}{15pt}
\renewcommand{\arraystretch}{2}

\begin{table*}[h!]
\centering
\begin{tabular}{|p{3cm}|c|c|}
\hline
\centering
Main chemical family & Chemical sub-family & SMARTS fragment \\
\hline \hline
\multirow{7}{4em}{\mbox{Saturated} \mbox{molecules}}
& Epoxy & [C;r3]-[O;r3]-[C;r3] \\
\cline{2-3}
& Aziridine & [C;r3]-[N;r3]-[C;r3] \\
\cline{2-3}
& Cyclopropane & [C;r3]-[C;r3]-[C;r3] \\
\cline{2-3}
& Oxetane / Azetidine & [C;r4]-[O,N;r4]-[C;r4] \\
\cline{2-3}
& Tetrahydrofurane / Pyrrolidine & [C;r5]-[O,N;r5]-[C;r5] \\
\cline{2-3}
& Ether / Alcohol & [C]-[O] \\
\cline{2-3}
& Other Aliphatic & [CX4] \\
\hline \hline
\multirow{5}{4em}{\mbox{Non-conjugated} \mbox{unsaturated} \mbox{molecules}}
& Ketone / Aldehyde & O=C \\
\cline{2-3}
& Cyano & N\#C \\
\cline{2-3}
& Alkyne & C\#C \\
\cline{2-3}
& Alkene & C=C \\
\cline{2-3}
& \renewcommand{\arraystretch}{1.5}
\begin{tabular}[c]{@{}c@{}} With no instances in the dataset: \\
Nitroso, Imine, Azo \end{tabular}
& O=N \,,\quad N=C \,,\quad N=N \\
\hline \hline
\multirow{5}{4em}{\mbox{$\pi$-conjugated} \mbox{molecules with} \mbox{3~conjugated} \mbox{atoms}}
& Amide & O=C-N \\
\cline{2-3}
& Carboximidate & N=C-O \\
\cline{2-3}
& Ester / Carboxylic Acid & O=C-O \\
\cline{2-3}
& Oxime & C=N-O \\
\cline{2-3}
& Amidine & N=C-N \\
\cline{2-3}
& \renewcommand{\arraystretch}{1.5}
\begin{tabular}[c]{@{}c@{}} With no instances in the dataset: \\
Azoxy, Enamine, Enol \\
Nitro / Nitrite, Diazo, Azide \ Isocyanate \end{tabular}
& \renewcommand{\arraystretch}{1.5}
\begin{tabular}[c]{@{}c@{}} N=N-O \,,\quad C=C-N \,,\quad C=C-O \\
O=N-O \,,\quad N=N-C \,,\quad N=N=N \end{tabular} \\
\hline \hline
\end{tabular}
\caption{This table shows the SMARTS fragments that were employed to recognise chemical sub-families within each of the main principal chemical families. These additional molecular descriptors were applied in the order in which they appear in the table, from top to bottom. We made no distinction of sub-families within the family of $\pi$-conjugated molecules with more than 3~conjugated atoms.}
\end{table*}

\begin{table*}[h!]
\centering
\begin{tabular}{c|c|c|c|c|c}
\cline{2-5}
& GTB2/SV \cite{Boleininger2016} & GTB2/SVP \cite{Boleininger2016} & DFT:PBE/cc-pVQZ \cite{Boleininger2016} & Expt. \cite{Olney1997} &  \\
\cline{1-5} \cline{2-5}
\multicolumn{1}{|l|}{H$_2$}   & 0.18 & 0.90 & 0.70 & 0.79 &  \\ \cline{1-5}
\multicolumn{1}{|l|}{methane} & 0.77 & 2.30 & 2.46 & 2.45 &  \\ \cline{1-5}
\multicolumn{1}{|l|}{ethane}  & 1.46 & 3.94 & 4.32 & 4.23 &  \\ \cline{1-5}
\multicolumn{1}{|l|}{propane} & 2.13 & 5.56 & 6.23 & 5.92 &  \\ \cline{1-5}
\multicolumn{1}{|l|}{butane}  & 2.81 & 7.22 & 8.14 & 7.69 &  \\ \cline{1-5}
\end{tabular}
\caption{This table shows an ordered subset of results obtained by Boleininger \textit{et al.} \cite{Boleininger2016}, exactly as they appear in their original publication. They correspond to mean polarizability volumes ($\alpha_{m}$) in Å$^{3}$, computed with the Gaussian polarizable-ion Tight Binding method with a second-order expansion of the charge density (GTB2) for H$_2$ and four saturated molecules (the first four elements in the homologous series of straight-chain alkanes). Calculations were carried out with a minimal basis set (SV) and a polarizable basis set (SVP). For comparison, we included experimental determinations by Olney \textit{et al.} \cite{Olney1997} and results from DFT-PBE with the correlation-consistent quadruple-zeta valence basis set (cc-pVQZ) \cite{Boleininger2016}.}
\end{table*}

\setlength{\tabcolsep}{2pt}
\renewcommand{\arraystretch}{1.3}
\renewcommand{\tablename}{\textbf{Table}}
\renewcommand{\thetable}{\textbf{III}}
\begin{table*}[h!]
\centering
\begin{tabular}{c|cccc|}
\cline{2-5} \cline{2-5}
\multirow{2}{*}{\textbf{{[}Polarizabilities{]}}} & \multicolumn{4}{c|}{\textbf{n: number of non-H atoms (C, O, N)}} \\
\cline{2-5} & \multicolumn{1}{c|}{\textbf{$n=2$}} & \multicolumn{1}{c|}{\textbf{$n=3$}} & \multicolumn{1}{c|}{\textbf{$n=4$}} & \textbf{$n=5$} \\
\hline \hline
\multicolumn{1}{|c|}{\begin{tabular}[c]{@{}c@{}}\textbf{$\cdot$C-C$\cdot$}\\ (alkanes)\end{tabular}} & \multicolumn{1}{c|}{\begin{tabular}[c]{@{}c@{}}$CH_3CH_3$\\ (ethane)\\ {[}4.226 Å$^3${]}\end{tabular}} & \multicolumn{1}{c|}{\begin{tabular}[c]{@{}c@{}}$H(CH_2)_2CH_3$\\ (propane)\\ {[}5.921 Å$^3${]}\end{tabular}} & \multicolumn{1}{c|}{\begin{tabular}[c]{@{}c@{}}$H(CH_2)_3CH_3$\\ (butane)\\ {[}8.020 Å$^3${]}\end{tabular}} & \begin{tabular}[c]{@{}c@{}}$H(CH_2)_4CH_3$\\ (pentane)\\ {[}9.88 Å$^3${]}\end{tabular} \\
\hline
\multicolumn{1}{|c|}{\begin{tabular}[c]{@{}c@{}}\textbf{$\cdot$C=C$\cdot$}\\ (alkenes)\end{tabular}} & \multicolumn{1}{c|}{\begin{tabular}[c]{@{}c@{}}$CH_2CH_2$\\ (ethylene)\\ {[}4.076 Å$^3${]}\end{tabular}} & \multicolumn{1}{c|}{\begin{tabular}[c]{@{}c@{}}$CH_3CHCH_2$\\ (propene)\\ {[}5.990 Å$^3${]}\end{tabular}} & \multicolumn{1}{c|}{\begin{tabular}[c]{@{}c@{}}$H(CH_2)_2CHCH_2$\\ (1-butene)\\ {[}7.830 Å$^3${]}\end{tabular}} & \begin{tabular}[c]{@{}c@{}}$H(CH_2)_3CHCH_2$\\ (1-pentene)\\ {[}9.65 Å$^3${]}\end{tabular} \\
\hline
\multicolumn{1}{|c|}{\begin{tabular}[c]{@{}c@{}}\textbf{$\cdot$C$\equiv$C$\cdot$}\\ (alkynes)\end{tabular}} & \multicolumn{1}{c|}{\begin{tabular}[c]{@{}c@{}}$CHCH$\\ (acetylene)\\ {[}3.487 Å$^3${]}\end{tabular}} & \multicolumn{1}{c|}{\begin{tabular}[c]{@{}c@{}}$CH_3CCH$\\ (propyne)\\ {[}5.550 Å$^3${]}\end{tabular}} & \multicolumn{1}{c|}{\begin{tabular}[c]{@{}c@{}}$H(CH_2)_2CCH$\\ (1-butyne)\\ {[}7.410 Å$^3${]}\end{tabular}} & \begin{tabular}[c]{@{}c@{}}$H(CH_2)_3CCH$\\ (1-pentyne)\\ {[}9.12 Å$^3${]}\end{tabular} \\
\hline
\multicolumn{1}{|c|}{\begin{tabular}[c]{@{}c@{}}$\cdot$\textbf{C=O}\\ (aldehydes)\end{tabular}} & \multicolumn{1}{c|}{\begin{tabular}[c]{@{}c@{}}$CH_2O$\\ (formaldehyde)\\ {[}2.770 Å$^3${]}\end{tabular}} & \multicolumn{1}{c|}{\begin{tabular}[c]{@{}c@{}}$CH_3CHO$\\ (acetaldehyde)\\ {[}4.278 Å$^3${]}\end{tabular}} & \multicolumn{1}{c|}{\begin{tabular}[c]{@{}c@{}}$H(CH_2)_2CHO$\\ (propanal)\\ {[}6.350 Å$^3${]}\end{tabular}} & \begin{tabular}[c]{@{}c@{}}$H(CH_2)_3CHO$\\ (butanal)\\ {[}8.20 Å$^3${]}\end{tabular} \\
\hline
\multicolumn{1}{|c|}{\begin{tabular}[c]{@{}c@{}}\textbf{$\cdot$C$\equiv$N}\\ (cyanides)\end{tabular}} & \multicolumn{1}{c|}{\begin{tabular}[c]{@{}c@{}}$HCN$\\ (hydrogen cyanide)\\ {[}2.346 Å$^3${]}\end{tabular}} & \multicolumn{1}{c|}{\begin{tabular}[c]{@{}c@{}}$CH_3CN$\\ (acetonitrile)\\ {[}4.280 Å$^3${]}\end{tabular}} & \multicolumn{1}{c|}{\begin{tabular}[c]{@{}c@{}}$H(CH_2)_2CN$\\ (propionitrile)\\ {[}6.240 Å$^3${]}\end{tabular}} & \begin{tabular}[c]{@{}c@{}}$H(CH_2)_3CN$\\ (butanenitrile)\\ {[}8.40 Å$^3${]}\end{tabular} \\
\hline \hline
\end{tabular}
\caption{This table shows an ordered subset of results that originally appeared in publications by Gussoni \textit{et al.} \cite{Gussoni1998} (also available at the NIST database \cite{NIST}) and Zevatskii \textit{et al.} \cite{Zevatskii2006}. The values correspond to experimentally measured electric dipole polarisabilities in Å$^{3}$ units, for non-conjugated unsaturated linear organic molecules containing the functional group at one end; we added linear alkanes for comparison. Note that for molecules of the same length (equal number of non-H atoms, $n$), the polarisabilities of alkenes are higher than those of the other unsaturated non-conjugated compounds; interestingly, a general trend can be extracted for these polarisability ($\alpha$) values: $\alpha$(alkanes) $>$ $\alpha$(alkenes) $>$ $\alpha$(alkynes) $>$ $\alpha$(aldehydes) $>$ $\alpha$(cyanides).}
\end{table*}

\setlength{\tabcolsep}{2pt}
\renewcommand{\arraystretch}{1.3}
\renewcommand{\tablename}{\textbf{Table}}
\renewcommand{\thetable}{\textbf{IV}}
\begin{table*}
\centering
\begin{tabular}{|c|c|c|c|c|}
\hline \hline
\textbf{\begin{tabular}[c]{@{}c@{}}$|\Delta_{CC2} f_{1}| < 0.01$\\ ( $|\Delta_{DFT} f_{1}| < 0.01$ )\end{tabular}} & \textbf{\begin{tabular}[c]{@{}c@{}}TD-DFT:PBE0\\ /def2SVP\end{tabular}} & \textbf{\begin{tabular}[c]{@{}c@{}}DFTB-Casida\\ :3OB\end{tabular}} & \textbf{\begin{tabular}[c]{@{}c@{}}DFTB-Casida\\ +LC:OB2\end{tabular}} & \textbf{\begin{tabular}[c]{@{}c@{}}DFTB-Casida\\ :3OB(H*)\end{tabular}} \\
\hline \hline
\textbf{All Saturated [5531]} & \textbf{87.7\%} & \textbf{\begin{tabular}[c]{@{}c@{}}45.2\%\\ (49.3\%)\end{tabular}} & \textbf{\begin{tabular}[c]{@{}c@{}}32.4\%\\ (34.9\%)\end{tabular}} & \textbf{\begin{tabular}[c]{@{}c@{}}46.7\%\\ (51.0\%)\end{tabular}} \\
\hline
$\hookrightarrow$ Epoxy [734] & 78.3\% & \begin{tabular}[c]{@{}c@{}}60.2\%\\ (72.8\%)\end{tabular} & \begin{tabular}[c]{@{}c@{}}56.5\%\\ (71.5\%)\end{tabular} & \begin{tabular}[c]{@{}c@{}}60.6\%\\ (73.0\%)\end{tabular} \\
\hline
$\hookrightarrow$ Aziridine [1010] & 86.4\% & \begin{tabular}[c]{@{}c@{}}43.9\%\\ (47.0\%)\end{tabular} & \begin{tabular}[c]{@{}c@{}}24.1\%\\ (28.0\%)\end{tabular} & \begin{tabular}[c]{@{}c@{}}45.7\%\\ (48.3\%)\end{tabular} \\
\hline
$\hookrightarrow$ Cyclopropane [1693] & 92.1\% & \begin{tabular}[c]{@{}c@{}}24.1\%\\ (24.7\%)\end{tabular} & \begin{tabular}[c]{@{}c@{}}11.8\%\\ (11.1\%)\end{tabular} & \begin{tabular}[c]{@{}c@{}}20.4\%\\ (21.3\%)\end{tabular} \\
\hline
$\hookrightarrow$ Oxetane/Azetidine [755] & 85.8\% & \begin{tabular}[c]{@{}c@{}}68.0\%\\ (74.6\%)\end{tabular} & \begin{tabular}[c]{@{}c@{}}54.6\%\\ (55.6\%)\end{tabular} & \begin{tabular}[c]{@{}c@{}}73.3\%\\ (78.9\%)\end{tabular} \\
\hline
\begin{tabular}[c]{@{}c@{}}$\hookrightarrow$ Tetrahydrofurane\\ /Pyrrolidine [276] \end{tabular} & 94.2\% & \begin{tabular}[c]{@{}c@{}}60.5\%\\ (66.3\%)\end{tabular} & \begin{tabular}[c]{@{}c@{}}52.9\%\\ (55.6\%)\end{tabular} & \begin{tabular}[c]{@{}c@{}}59.4\%\\ (61.2\%)\end{tabular} \\
\hline
$\hookrightarrow$ Ether/Alcohol [852] & 90.6\% & \begin{tabular}[c]{@{}c@{}}52.8\%\\ (52.3\%)\end{tabular} & \begin{tabular}[c]{@{}c@{}}39.7\%\\ (36.5\%)\end{tabular} & \begin{tabular}[c]{@{}c@{}}59.6\%\\ (64.7\%)\end{tabular} \\
\hline
$\hookrightarrow$ Other Aliphatic [211] & 78.0\% & \begin{tabular}[c]{@{}c@{}}36.8\%\\ (40.2\%)\end{tabular} & \begin{tabular}[c]{@{}c@{}}19.6\%\\ (19.6\%)\end{tabular} & \begin{tabular}[c]{@{}c@{}}50.7\%\\ (56.0\%)\end{tabular} \\
\hline \hline
\textbf{\begin{tabular}[c]{@{}c@{}}All Unsaturated\\ non-conjugated [6757] \end{tabular}} & \textbf{86.1\%} & \textbf{\begin{tabular}[c]{@{}c@{}}72.3\%\\ (77.8\%)\end{tabular}}  & \textbf{\begin{tabular}[c]{@{}c@{}}73.7\%\\ (78.1\%)\end{tabular}} & \textbf{\begin{tabular}[c]{@{}c@{}}72.4\%\\ (78.2\%)\end{tabular}} \\
\hline
$\hookrightarrow$ Ketone/Aldehyde [2775] & 100.0\% & \begin{tabular}[c]{@{}c@{}}96.0\%\\ (96.2\%)\end{tabular} & \begin{tabular}[c]{@{}c@{}}99.3\%\\ (99.3\%)\end{tabular} & \begin{tabular}[c]{@{}c@{}}95.6\%\\ (96.0\%)\end{tabular} \\
\hline
$\hookrightarrow$ Alkyne [1520] & 86.5\% & \begin{tabular}[c]{@{}c@{}}69.9\%\\ (76.4\%)\end{tabular} & \begin{tabular}[c]{@{}c@{}}65.1\%\\ (69.3\%)\end{tabular} & \begin{tabular}[c]{@{}c@{}}72.4\%\\ (79.9\%)\end{tabular} \\
\hline
$\hookrightarrow$ Cyano [1443] & 75.1\% & \begin{tabular}[c]{@{}c@{}}56.4\%\\ (70.8\%)\end{tabular} & \begin{tabular}[c]{@{}c@{}}55.2\%\\ (66.3\%)\end{tabular} & \begin{tabular}[c]{@{}c@{}}55.2\%\\ (69.3\%)\end{tabular} \\
\hline
$\hookrightarrow$ Alkene [1019] & 63.6\% & \begin{tabular}[c]{@{}c@{}}33.7\%\\ (39.6\%)\end{tabular} & \begin{tabular}[c]{@{}c@{}}42.8\%\\ (50.2\%)\end{tabular} & \begin{tabular}[c]{@{}c@{}}33.3\%\\ (40.7\%)\end{tabular} \\
\hline \hline
\end{tabular}
\caption{This table shows the percentage of molecules with oscillator strengths ($f_1$) that are considered to be good estimates (compared to CC2), for TD-DFT:PBE0/def2SVP and the DFTB-approximate ES methods of Fig.~S7 (in this ESI). In other words, this table shows for each chemical family and subgroup the percentage of molecules with absolute $f_1$ prediction errors ($|\Delta_{CC2} f_1|$) that fell below the 0.01 threshold (depicted as dashed lines in Fig.~S7), a value that is customarily used to distinguish between dark ($f < 0.01$) and bright ($f >= 0.01$) excitations. In parentheses, we report the percentage of molecules with oscillator strengths that are considered to be similar between the compared methods (i.e., $|\Delta_{DFT} f_1| < 0.01$). In the first column, for each chemical family and subgroup we also indicate the number of occurrences in the data set (inside the square brackets). Here we show data for saturated and non-conjugated unsaturated molecules; please refer to Table~V within this ESI for data corresponding to $\pi$-conjugated molecules.}
\end{table*}

\setlength{\tabcolsep}{2pt}
\renewcommand{\arraystretch}{1.3}
\renewcommand{\tablename}{\textbf{Table}}
\renewcommand{\thetable}{\textbf{V}}
\begin{table*}
\centering
\begin{tabular}{|c|c|c|c|c|}
\hline \hline
\textbf{\begin{tabular}[c]{@{}c@{}}$|\Delta_{CC2} f_{1}| < 0.01$\\ ( $|\Delta_{DFT} f_{1}| < 0.01$ )\end{tabular}} & \textbf{\begin{tabular}[c]{@{}c@{}}TD-DFT:PBE0\\ /def2SVP\end{tabular}} & \textbf{\begin{tabular}[c]{@{}c@{}}DFTB-Casida\\ :3OB\end{tabular}} & \textbf{\begin{tabular}[c]{@{}c@{}}DFTB-Casida\\ +LC:OB2\end{tabular}} & \textbf{\begin{tabular}[c]{@{}c@{}}DFTB-Casida\\ :3OB(H*)\end{tabular}} \\
\hline \hline
\textbf{\begin{tabular}[c]{@{}c@{}}Unsaturated $\pi$-conjugated\\ with 3 conj. atoms [3166] \end{tabular}} & \textbf{94.8\%} & \textbf{\begin{tabular}[c]{@{}c@{}}88.3\%\\ (91.4\%)\end{tabular}}  & \textbf{\begin{tabular}[c]{@{}c@{}}88.4\%\\ (92.3\%)\end{tabular}} & \textbf{\begin{tabular}[c]{@{}c@{}}89.2\%\\ (91.5\%)\end{tabular}} \\
\hline
$\hookrightarrow$ Amide [1186] & 99.0\% & \begin{tabular}[c]{@{}c@{}}95.3\%\\ (96.1\%)\end{tabular} & \begin{tabular}[c]{@{}c@{}}97.2\%\\ (98.5\%)\end{tabular} & \begin{tabular}[c]{@{}c@{}}96.4\%\\ (96.4\%)\end{tabular} \\
\hline
$\hookrightarrow$ Carboximidate [1000] & 93.4\% & \begin{tabular}[c]{@{}c@{}}80.4\%\\ (85.3\%)\end{tabular} & \begin{tabular}[c]{@{}c@{}}78.2\%\\ (85.5\%)\end{tabular} & \begin{tabular}[c]{@{}c@{}}81.7\%\\ (84.6\%)\end{tabular} \\
\hline
$\hookrightarrow$ Ester/Carboxylic Acid [678] & 98.4\% & \begin{tabular}[c]{@{}c@{}}95.9\%\\ (95.7\%)\end{tabular} & \begin{tabular}[c]{@{}c@{}}97.05\%\\ (96.5\%)\end{tabular} & \begin{tabular}[c]{@{}c@{}}96.0\%\\ (95.7\%)\end{tabular} \\
\hline
$\hookrightarrow$ Oxime [200] & 81.0\% & \begin{tabular}[c]{@{}c@{}}76.0\%\\ (83.0\%)\end{tabular} & \begin{tabular}[c]{@{}c@{}}78.0\%\\ (83.5\%)\end{tabular} & \begin{tabular}[c]{@{}c@{}}75.5\%\\ (86.0\%)\end{tabular} \\
\hline
$\hookrightarrow$ Amidine [102] & 61.8\% & \begin{tabular}[c]{@{}c@{}}58.8\%\\ (84.3\%)\end{tabular} & \begin{tabular}[c]{@{}c@{}}50.0\%\\ (76.5\%)\end{tabular} & \begin{tabular}[c]{@{}c@{}}59.8\%\\ (84.3\%)\end{tabular} \\
\hline \hline
\textbf{\begin{tabular}[c]{@{}c@{}}Unsaturated $\pi$-conjugated\\ with $>$3 conj. atoms [6332] \end{tabular}} & \textbf{65.0\%} & \textbf{\begin{tabular}[c]{@{}c@{}}54.9\%\\ (64.2\%)\end{tabular}}  & \textbf{\begin{tabular}[c]{@{}c@{}}51.0\%\\ (58.7\%)\end{tabular}} & \textbf{\begin{tabular}[c]{@{}c@{}}54.9\%\\ (63.3\%)\end{tabular}} \\
\hline
$\hookrightarrow$ 4 Conj. atoms [965] & 87.8\% & \begin{tabular}[c]{@{}c@{}}82.0\%\\ (87.7\%)\end{tabular} & \begin{tabular}[c]{@{}c@{}}83.5\%\\ (88.0\%)\end{tabular} & \begin{tabular}[c]{@{}c@{}}82.8\%\\ (88.5\%)\end{tabular} \\
\hline
$\hookrightarrow$ 5 Conj. atoms [1552] & 68.1\% & \begin{tabular}[c]{@{}c@{}}63.1\%\\ (70.1\%)\end{tabular} & \begin{tabular}[c]{@{}c@{}}62.8\%\\ (69.3\%)\end{tabular} & \begin{tabular}[c]{@{}c@{}}64.7\%\\ (68.5\%)\end{tabular} \\
\hline
$\hookrightarrow$ 6 Conj. atoms [1357] & 58.2\% & \begin{tabular}[c]{@{}c@{}}42.9\%\\ (49.7\%)\end{tabular} & \begin{tabular}[c]{@{}c@{}}34.2\%\\ (41.1\%)\end{tabular} & \begin{tabular}[c]{@{}c@{}}42.5\%\\ (48.7\%)\end{tabular} \\
\hline
$\hookrightarrow$ 7 Conj. atoms [1457] & 59.8\% & \begin{tabular}[c]{@{}c@{}}46.7\%\\ (59.6\%)\end{tabular} & \begin{tabular}[c]{@{}c@{}}38.3\%\\ (48.7\%)\end{tabular} & \begin{tabular}[c]{@{}c@{}}45.4\%\\ (58.5\%)\end{tabular} \\
\hline
$\hookrightarrow$ 8 Conj. atoms [1001] & 55.0\% & \begin{tabular}[c]{@{}c@{}}44.3\%\\ (58.8\%)\end{tabular} & \begin{tabular}[c]{@{}c@{}}42.5\%\\ (52.5\%)\end{tabular} & \begin{tabular}[c]{@{}c@{}}43.5\%\\ (57.4\%)\end{tabular} \\
\hline \hline
\end{tabular}
\caption{This table shows the percentage of molecules with oscillator strengths ($f_1$) that are considered good estimates (compared to CC2), for TD-DFT:PBE0/def2SVP and DFTB-approximate ES methods in Fig.~S5 (in this ESI). In other words, this table shows for each chemical family and subgroup the percentage of molecules with absolute $f_1$ prediction errors ($|\Delta_{CC2} f_1|$) that fell below the 0.01 threshold (depicted as dashed lines in Fig.~S5), a value that is customarily used to distinguish between dark ($f < 0.01$) and bright ($f >= 0.01$) excitations. In parentheses, we report the percentage of molecules with oscillator strengths that are considered to be similar between the compared methods (i.e., $|\Delta_{DFT} f_1| < 0.01$). In the first column, for each chemical family and subgroup we also indicate the number of occurrences in the data set (inside the square brackets). Here we show data for $\pi$-conjugated molecules; please refer to Table~IV within this ESI for data corresponding to saturated and non-conjugated unsaturated molecules.}
\end{table*}

\renewcommand{\figurename}{\textbf{Figure S1}}
\begin{figure*}
\includegraphics[scale=0.6]{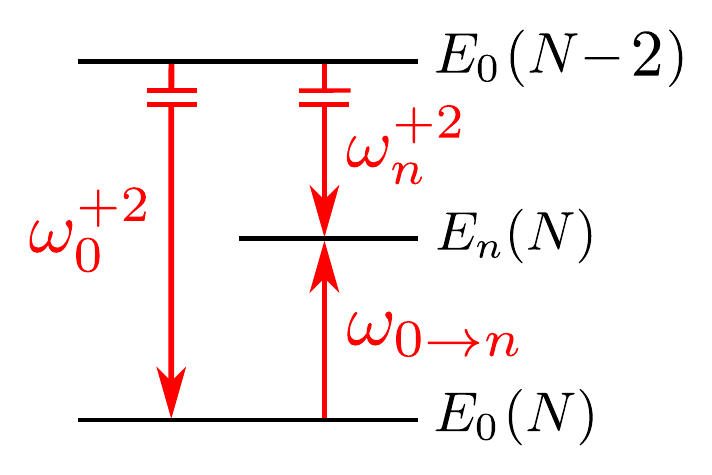}
\caption{\label{FigureS1} Diagram depicting how ppRPA computes excitation energies from the energies of two-electron addition processes. An excitation energy for the N-electron system ($\omega_{0 \rightarrow n}$) can be computed as the difference between the energies of two-electron addition processes ($\omega_{0}^{+2}$ and $\omega_{n}^{+2}$), from the ground state of a two-electron deficient system (i.e. with N-2 electrons) into the ground and $n$-th excited states of the resulting N-electron system. The horizontal bars represent electronic states of the same system, with energies $E_{0}\left( N-2 \right) >> E_{n}\left( N \right) > E_{0}\left( N \right)$.}
\end{figure*}

\renewcommand{\figurename}{\textbf{Figure S2}}
\begin{figure*}
\includegraphics[scale=0.6]{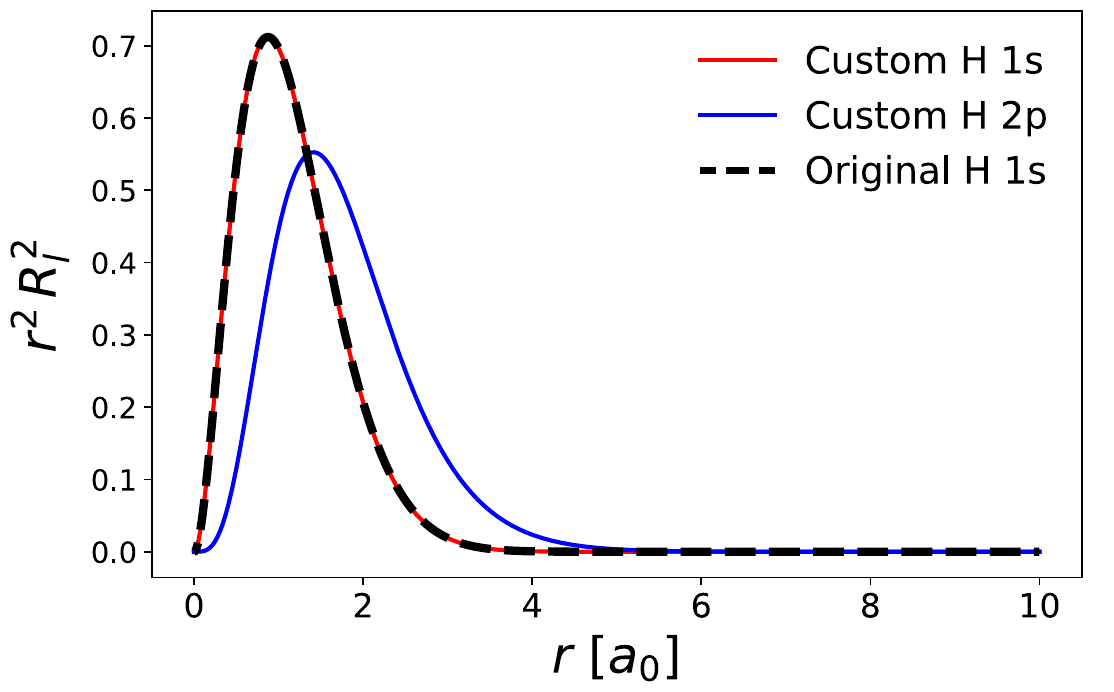}
\caption{\label{FigureS2} Radial probability densities for the hydrogen orbitals in the original 3OB and in the custom H-polarized 3OB(H*) parameter sets. As was intended, the custom $1s$~orbital for H in the 3OB(H*) parameter set (red solid line) coincides with that of the original 3OB set (black dashed line). The three $2p$~orbitals in 3OB(H*), allowing polarization, have a radial wave function with a probability density (blue solid line) that partially overlaps that of the $1s$~orbital.}
\end{figure*}

\renewcommand{\figurename}{\textbf{Figure S3}}
\begin{figure*}[h!]
\includegraphics[scale=0.625]{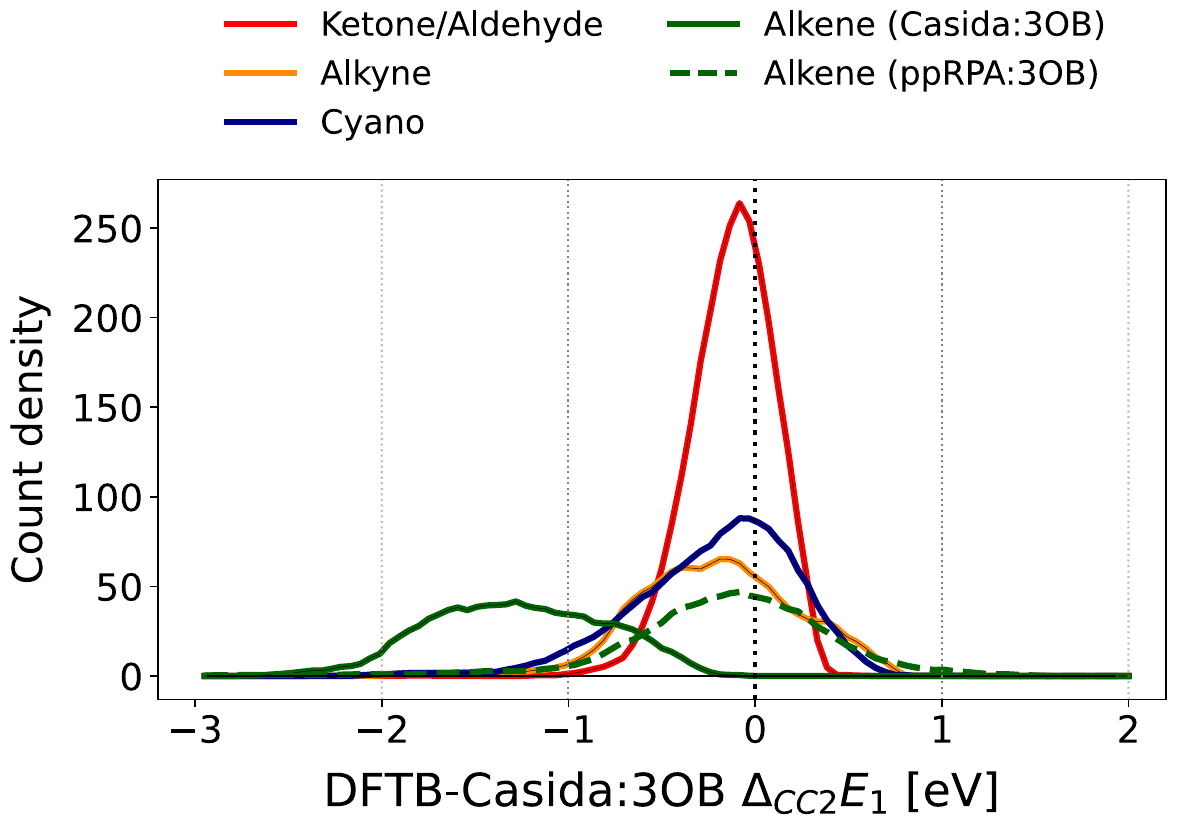}
\label{FigureS3}
\caption{Overlapped $E_1$ error distributions for chemical subfamilies within the family of non-conjugated unsaturated molecules (i.e. molecules characterised by isolated double and triple bonds). Histograms were computed using results from DFTB-Casida:3OB (solid lines) and DFTB-ppRPA:3OB (dashed line). See the first column of Table~IV, in this ESI, for the number of occurrences per chemical family and subgroup.}
\end{figure*}

\renewcommand{\figurename}{\textbf{Figure S4}}
\begin{figure*}[h!]
\includegraphics[scale=0.625]{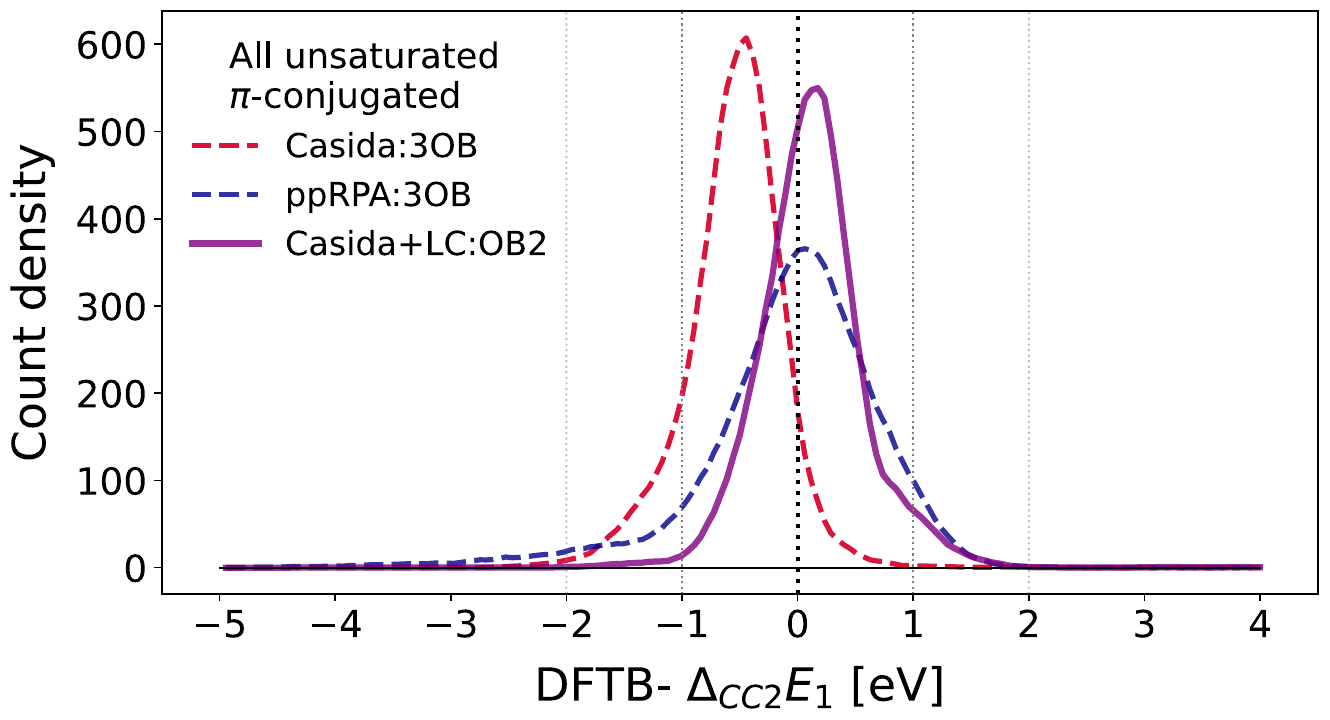}
\caption{\label{FigureS4} Overlapped $E_1$ error distributions for all the unsaturated $\pi$-conjugated molecules, computed using DFTB-Casida:3OB (red dashed line), DFTB-ppRPA:3OB (blue dashed line) and DFTB-Casida+LC:3OB (purple solid line). Note that including long-range corrections into DFTB-Casida computations (i.e. computing with DFTB+LC-Casida:OB2) achieves an accuracy similar to that of DFTB-ppRPA:3OB, with a slightly better precision. See the first column of Table~V, in this ESI, for the number of occurrences per chemical family and subgroup.}
\end{figure*}

\renewcommand{\figurename}{\textbf{Figure S5}}
\begin{figure*}[h!]
\includegraphics[scale=0.625]{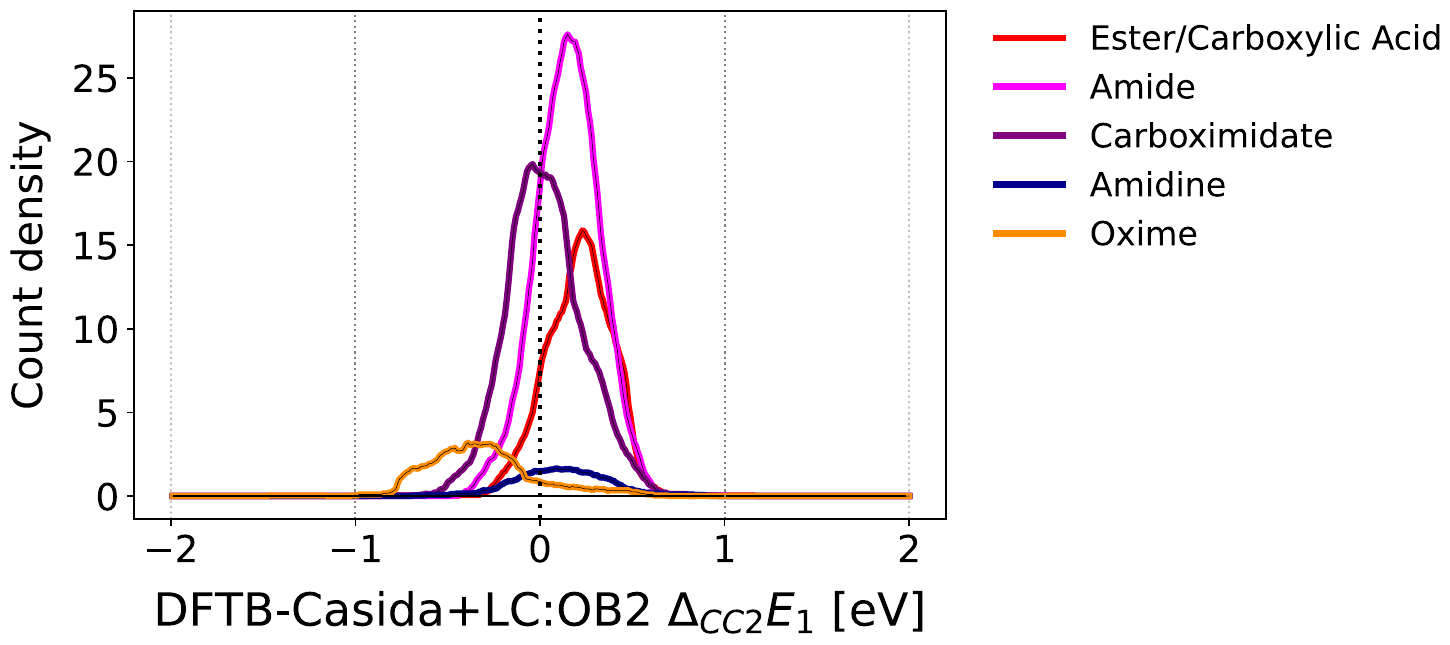}
\label{FigureS5}
\caption{Overlapped $E_1$ error distributions for chemical subfamilies within the family of unsaturated $\pi$-conjugated molecules with 3~conjugated atoms. Histograms were computed using results from DFTB-Casida+LC:OB2. See the first column of Table~V, in this ESI, for the number of occurrences per chemical family and subgroup.}
\end{figure*}

\renewcommand{\figurename}{\textbf{Figure S6}}
\begin{figure*}[h!]
\includegraphics[scale=0.6]{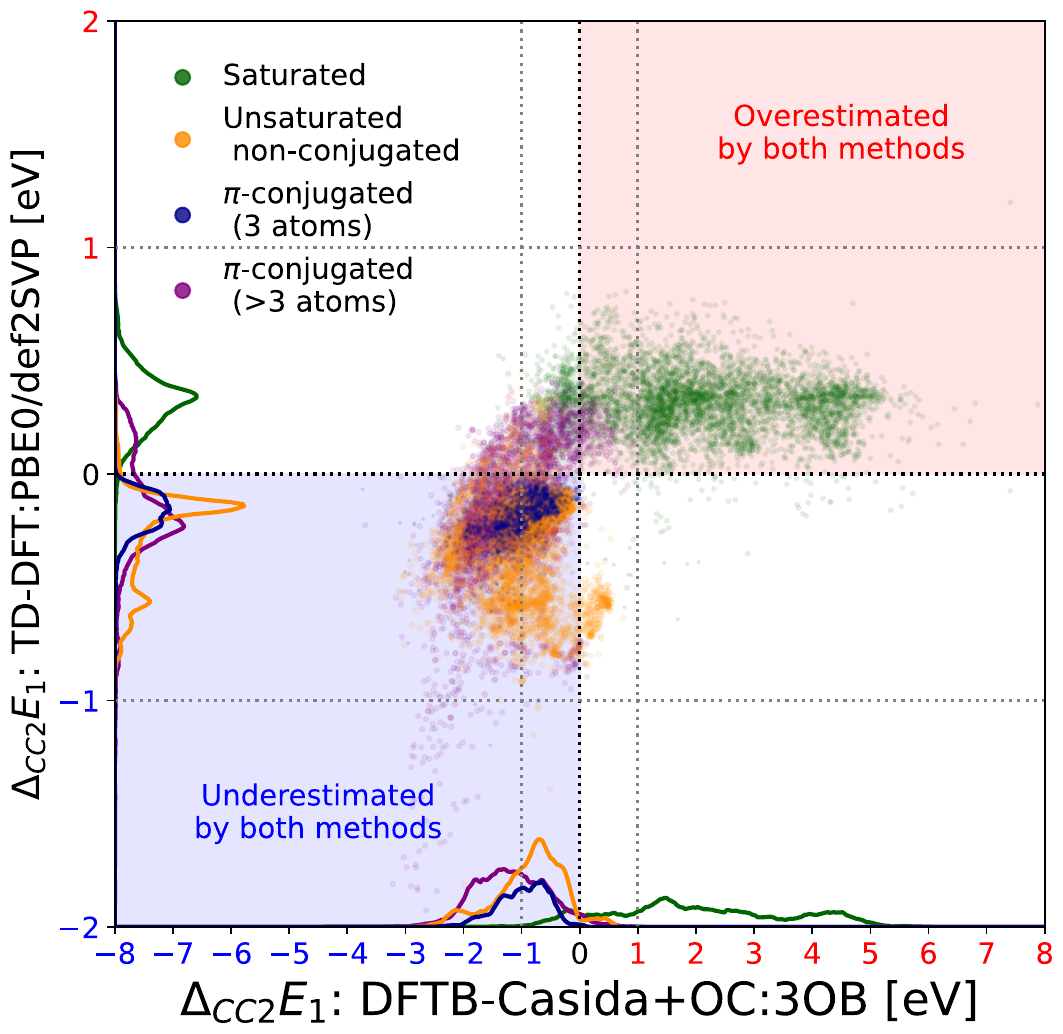}
\caption{\label{FigureS6} Comparison of prediction errors $\Delta_{CC2} E_1$ for TD-DFT:PBE0/def2SVP and DFTB-Casida+OC:3OB. The scattered datapoints correspond to each of the nearly 21,800 molecules in the GDB-8 chemical subspace. The datapoints and the projected histograms were coloured according to the main chemical identity of the compounds: green for saturated molecules, orange for non-conjugated molecules with an isolated double or triple bond, and blue and purple for $\pi$-conjugated molecules with~3 or more conjugated atoms, respectively. See the first column of Tables~IV~\&~V, in this ESI, for the number of occurrences per chemical family and subgroup.}
\end{figure*}

\renewcommand{\figurename}{\textbf{Figure S7}}
\begin{figure*}[h!]
\includegraphics[scale=0.625]{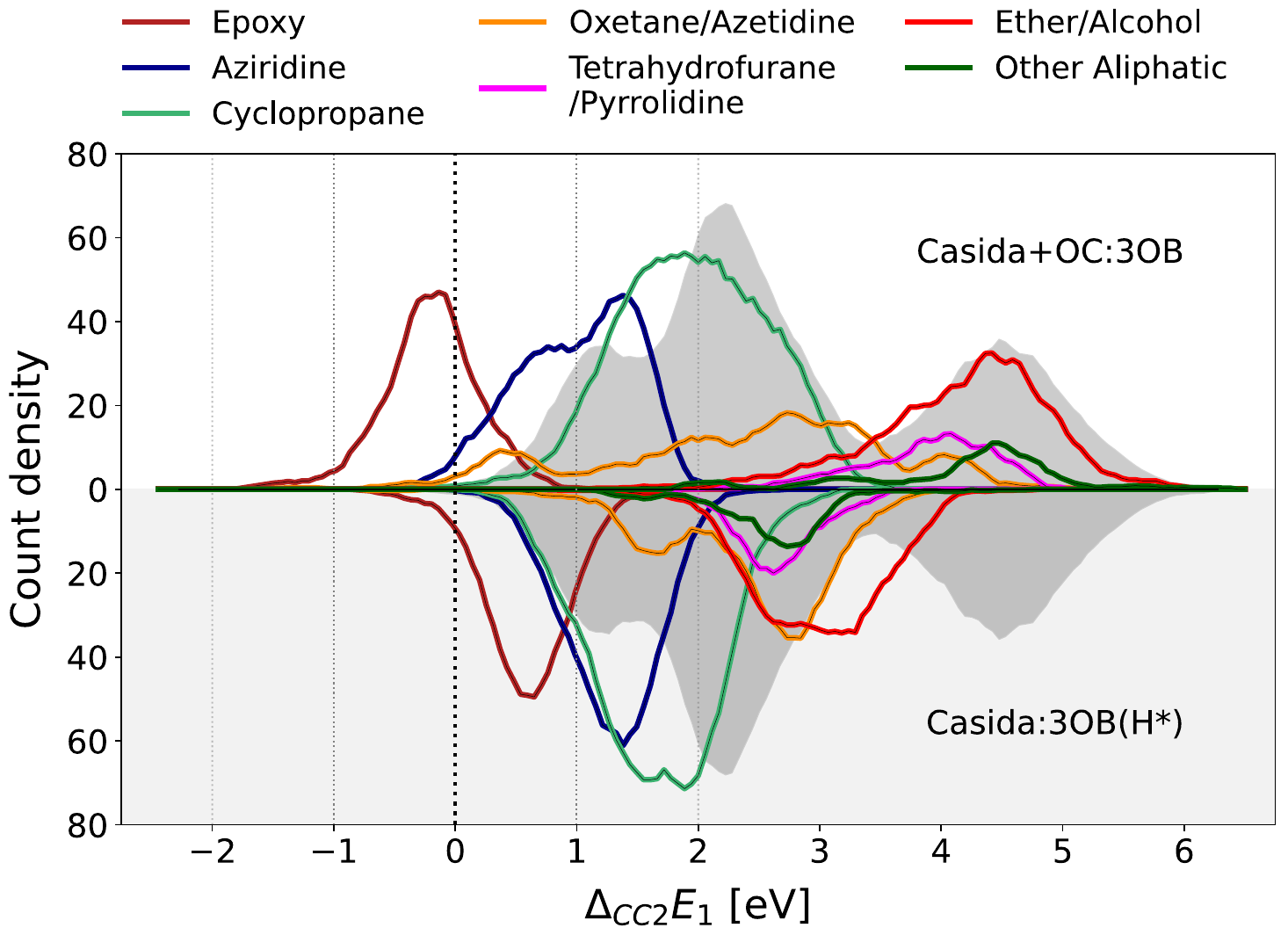}
\caption{\label{FigureS7} Overlapped $E_1$ error distributions for chemical groups within the family of saturated molecules. On-site corrected TD-DFTB results, obtained with DFTB-Casida+OC:3OB, are displayed at the top, on a white background. Histograms computed from results obtained with the minimally polarized DFTB-Casida:3OB(H*) are shown at the bottom, on a grey background. For ease of comparison, in both backgrounds we have included to scale, as dark-grey shaded areas, the original $E_1$ error distribution for DFTB-Casida:3OB corresponding to all the saturated molecules. See the first column of Tables~IV~\&~V, in this ESI, for the number of occurrences per chemical family and subgroup.}
\end{figure*}

\renewcommand{\figurename}{\textbf{Figure S8}}
\begin{figure*}
\includegraphics[scale=0.425]{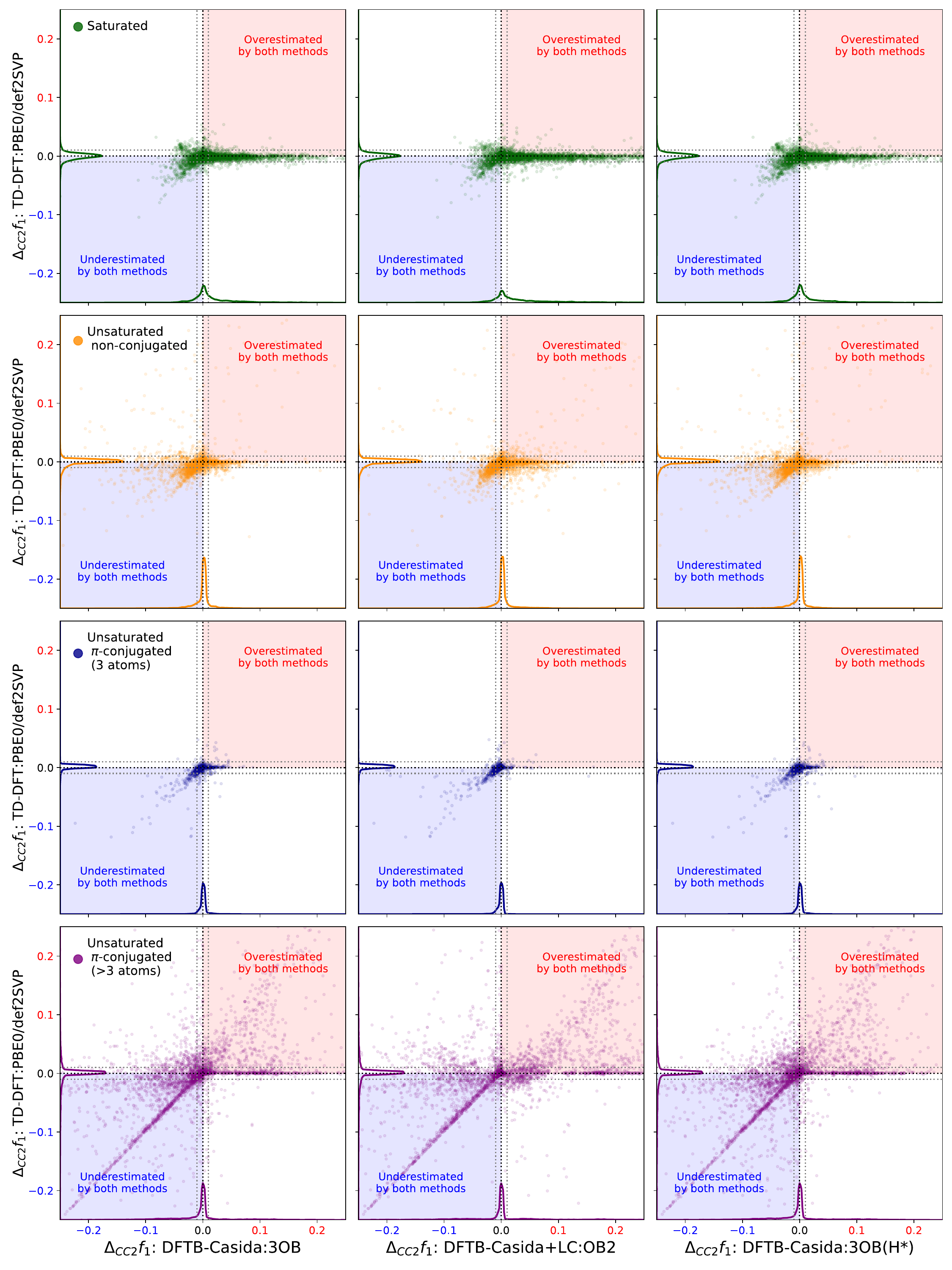}
\caption{\label{FigureS8} Comparison of prediction errors $\Delta_{CC2} f_1$, where $f_1$ is the oscillator strength associated to $E_1$, for TD-DFT:PBE0/def2SVP and three different DFTB-approximate approaches: Casida:3OB \textit{(1st~column)}, Casida+LC:OB2 \textit{(2nd~column)} and Casida:3OB(H*) \textit{(3rd~column)}. The scattered datapoints correspond to each of the nearly 21,800 molecules in the GDB-8 chemical subspace. The datapoints and the projected histograms were coloured according to the main chemical identity of the compounds: green for saturated molecules \textit{(1st~row)}, orange for non-conjugated molecules with an isolated double or triple bond \textit{(2nd~row)}, and blue and purple for $\pi$-conjugated molecules with~3 or more conjugated atoms \textit{(3rd~and 4th~rows, respectively)}. See the first column of Tables~IV~\&~V, in this ESI, for the number of occurrences per chemical family and subgroup.}
\end{figure*}